\documentclass[11pt,a4paper]{article}
\usepackage[utf8]{inputenc}
\usepackage{jheppub}
\usepackage{mathrsfs}
\usepackage{graphicx}

\allowdisplaybreaks

\def\Res_#1{\operatorname*{Res}_{#1}}

\def\nn{\nonumber}

\def\braket#1{\langle #1 \rangle}
\def\bbraket#1{\langle\!\langle #1 \rangle\!\rangle}

\def\ie{i.e. }
\def\eg{e.g. }
\def\etc{etc. }
\def\cf{cf. }
\def\eqn#1{eq.~\eqref{#1}}
\def\eqns#1#2{eqs.~\eqref{#1} and~\eqref{#2}}

\def\Eqn#1{Eq.~\eqref{#1}}

\def\fig#1{figure~{\ref{#1}}}

\def\sec#1{section~{\ref{#1}}}

\def\app#1{appendix~{\ref{#1}}}

\title{Twisted de Rham theory for string double copy in AdS}
\author[1]{Hiren Kakkad,}
\author[1]{Alexander Ochirov}
\author[1]{and Shijie Zhang}

\affiliation[1]{Center for Fundamental Physics, School of Physical Science and Technology, \\
ShanghaiTech University, 393 Middle Huaxia Road, Shanghai 201210, China}

\emailAdd{\{hkakkad,ochirov,zhangshj2025\}@shanghaitech.edu.cn}

\abstract{This work is motivated by the recent evidence for a double-copy relationship between open- and closed-string amplitudes in Anti-de Sitter (AdS) space. At present, the evidence has the form of a double-copy relation for string-amplitude building blocks, which are combined using the multiple-polylogarithm (MPL) generating functions. These generate MPLs relevant for all-order AdS curvature corrections of four-point string amplitudes. In this paper, we prove this building-block double copy using a new, noncommutative version of twisted de Rham theory. In flat space, the usual twisted de Rham theory is already known to be a natural framework to describe the Kawai-Lewellen-Tye (KLT) double-copy map from open- to closed-string amplitudes, in which the KLT kernel can be computed from the intersections of the open-string amplitude integration contours. We formulate twisted de Rham theory for noncommutative-ring-valued differential forms on complex manifolds and use it to derive the intersection number of two open-string contours, which are closed in the noncommutative twisted homology sense. The inverse of this intersection number is precisely the AdS double-copy kernel for the four-point open- and closed-string generating functions.
}

\begin{document}
\maketitle
\addtocontents{toc}{\protect\setcounter{tocdepth}{2}}

%%%%%%%%%%%%%%%%%%%%%%%%%%%%%%%%%%%%%%%%%%%%%%%%%%
\section{Introduction}
\label{sec:intro}
%%%%%%%%%%%%%%%%%%%%%%%%%%%%%%%%%%%%%%%%%%%%%%%%%%

The fascinating network of connections between various quantum field theories~\cite{Bern:2008qj,Bern:2010ue,Cachazo:2013hca,Cachazo:2013iea,Johansson:2014zca,Chiodaroli:2015rdg,Johansson:2019dnu,Bautista:2019evw,Bern:2019prr}, which includes gravity as a ``double copy'' of gauge theory, started from the Kawai-Lewellen-Tye (KLT) relations~\cite{Kawai:1985xq} between open- and closed-string tree-level scattering amplitudes:
\begin{equation}
{\cal A}_n^\text{closed} = \sum_{\beta,\gamma} A_n^\text{open}(\beta)\,S_{\beta|\gamma}(\alpha')\,A_n^\text{open}(\gamma) ,
\label{eq:KLT}
\end{equation}
Here $\beta$ and $\gamma$ are the color orderings of the open-string partial amplitudes, and $S_{\beta|\gamma}(\alpha')$ is the KLT kernel.
It involves the external momenta and the inverse string tension $\alpha'$ and may be traced back to the monodromy structure of the open-string worldsheet integrals.
Gluons and graviton scattering may be obtained in the field-theory limit $\alpha' \to 0$, in which the KLT relations retain the schematic form~\cite{Bern:1998sv,BjerrumBohr:2009rd,BjerrumBohr:2010zs,Bjerrum-Bohr:2010diw,Bjerrum-Bohr:2010kyi,Bjerrum-Bohr:2010pnr} of \eqn{eq:KLT}.
In this paper, we are interested in understanding how to expand the scope of the string-theoretic KLT relations.
To introduce the framework that we believe will enable this, let us take a look at the simplest nontrivial setting of four-point tree-level amplitudes.
For the open string, after fixing three vertex-operator positions using the worldsheet ${\rm SL}(2,\mathbb{R})$ invariance, the amplitude reduces to an integral over a single real puncture location~$z$ in the interval~$(0,1)$.
The resulting amplitude for the scattering of four tachyons is the celebrated Veneziano amplitude~\cite{Veneziano:1968yb}, the color-ordered version of which is (proportional to)
\begin{equation}
\int_0^1\!z^{s-1} (1-z)^{t-1} dz = B(s,t) = \frac{\Gamma(s)\Gamma(t)}{\Gamma(s+t)} ,
\label{eq:VenezianoAmplitude}
\end{equation}
where $s =-\alpha'(p_1{+}p_2)^2$ and $t =-\alpha'(p_2{+}p_3)^2$ are string-coupling-rescaled Mandelstam invariants (or shifts thereof).
The corresponding fully symmetric Virasoro-Shapiro closed-string amplitude for four tachyons~\cite{Virasoro:1969me,Shapiro:1969km} is  (proportional to)
\begin{equation}
\int_{\mathbb{C}\setminus\{0,1\}}\!\!dz \wedge d\bar{z}\,|z|^{2s-2} |1-z|^{2t-2}
 = -\frac{2\pi i\,\Gamma(s)\Gamma(t)\Gamma(1-s-t)}{\Gamma(s+t)\Gamma(1-s)\Gamma(1-t)} .
\label{eq:VirasoroShapiroAmplitude}
\end{equation}
Here the kinematic identification involves an imaginary mass ($m^2=-1/\alpha'$):
\begin{equation}
s = -1-\alpha'(p_1+p_2)^2 , \qquad
t = -1-\alpha'(p_2+p_3)^2, \qquad
u = -1-\alpha'(p_1+p_3)^2 = 1-s-t .
\label{eq:Mandelstam4}
\end{equation}
Note that the identity $\sin(\pi x) = \pi/[\Gamma(x)\Gamma(1-x)]$ relates the two results~\eqref{eq:VenezianoAmplitude} and~\eqref{eq:VirasoroShapiroAmplitude}:
\begin{equation}
\frac{\pi\,\Gamma(s)\Gamma(t)\Gamma(1-s-t)}{\Gamma(s+t)\Gamma(1-s)\Gamma(1-t)}
 = \frac{\sin(\pi s) \sin(\pi t)}{\sin(\pi(s + t))} B(s,t)^2 ,
\label{eq:Veneziano2VirasoroShapiro}
\end{equation}
which is an explicit example of the KLT relations~\eqref{eq:KLT}.

\paragraph{Scattering amplitudes in Anti-de Sitter}\!\!\!\!(AdS) space differ qualitatively from their flat-space counterparts.
The nonzero curvature modifies their analytic structure, and amplitudes are more naturally interpreted in terms of boundary correlators in the dual conformal field theory.
Recently, a systematic framework for introducing AdS curvature corrections to the flat-space tree-level open- and closed-string amplitudes has been developed in \cite{Alday:2023kfm,Alday:2024yax,Alday:2024ksp} and \cite{Alday:2022uxp,Alday:2022xwz,Alday:2023jdk,Alday:2023mvu,Alday:2024rjs}, respectively.
A key insight is that these corrections can be decomposed into universal ``building blocks'' that generalize the four-point Veneziano and Virasoro-Shapiro integrals.
The curvature corrections to the Veneziano amplitude can be accounted for via introducing multiple polylogarithms (MPLs) $L_w(x)$ \cite{Chen:1977oja,Goncharov:1998kja,MINH2000217,Remiddi:1999ew,Goncharov:2001iea,Brown:2004ugm} in the integral
\begin{equation}
J_w(s,t)=\int_0^1 z^{s-1}(1-z)^{t-1}L_w(z) dz ,
\label{eq:open_string_BB}
\end{equation}
where the MPLs are enumerated by words~$w$ formed with letters in the alphabet~$\{0,1\}$.
For an empty word, the result reduces back to the flat-space amplitude~\eqref{eq:VenezianoAmplitude}.
This infinite tower of integrals is conveniently combined into a single object by introducing two \emph{non-commuting} variables $e_0$ and $e_1$, one for each letter.
This gives rise to the MPL generating function~\cite{MINH2000217}
\begin{equation}
\begin{aligned}
\label{eq:GenFunctionL}
L(e_0,e_1;x)&=1+L_0(x)e_0+L_1(x)e_1+L_{00}(x)e_0^2+L_{01}(x)e_0e_1+ \dots \\ &
 =1+\log (x)e_0+\log(1-x)e_1+\frac{1}{2}\log^2(x)e_0^2+{\rm Li}_2(x)[e_0,e_1]+\dots
\end{aligned}
\end{equation} 
A similar generating function for the curvature-induced open-string integrals~\eqref{eq:open_string_BB} is then
\begin{equation}
J(e_\ell;s,t)=\int_0^1\!z^{s-1}(1-z)^{t-1}L(e_\ell;z) dz , \quad \text{where} \quad \ell \in \{0,1\} .
\label{eq:open_string_AdS}
\end{equation}

For the (curvature-induced) closed-string integrals, on the other hand, one introduces single-valued multiple polylogarithms (SVMPL) ${\cal L}_w(z)$, which may similarly be collected into a generating function ${\cal L}(e_\ell;z)$.
The corresponding integral generating function reads
\begin{equation}
I(e_\ell;s,t)=\int_{\mathbb{C}\setminus\{0,1\}}\!|z|^{2s-2}|1-z|^{2t-2}{\cal L}(e_\ell;z) d^2z .
\label{eq:closed_string_AdS}
\end{equation}
An important point, which we will exploit in this paper, is that these generating series are related via a single-valued map~\cite{Brown:2004ugm}
\begin{equation}
{\cal L}(e_\ell;z) = L(e_\ell;z) L^{\rm R}(e'_\ell;\bar{z}) ,
\label{eq:SingleValuedMap}
\end{equation}
where the superscript ${\rm R}$ stands for ``reversal'', which we discuss in more detail in \sec{sec:dual}.
This relation already hints towards a possible double copy in AdS between the open-string integral~\eqref{eq:open_string_AdS} and the closed-string integral~\eqref{eq:closed_string_AdS}.
Indeed, Alday, Nocchi and Str\"omholm Sangar\'e~\cite{Alday:2025bjp} have explicitly demonstrated such a building-block double copy
\begin{equation}
I(e_\ell;s,t) = J(e_\ell;s,t)\,{\cal K}(e_\ell;s,t)\,J^{\rm R}(e'_\ell;s,t) .
\label{eq:AdS_double_copy}
\end{equation}
with the KLT kernel
\begin{equation}
{\cal K}(e_\ell;s,t)=\frac{i}{2}\bigg(1+\frac{e^{2i\pi s}M_0}{1-e^{2i\pi s}M_0}+\frac{e^{2i\pi t}M_1}{1-e^{2i\pi t }M_1}\bigg)^{\!-1} .
\end{equation}
depending only on the combined monodromy factors $e^{2i\pi s}M_0$ and $e^{2i\pi t}M_1$ at~0 and~1 of the Koba-Nielsen factor $z^s (1-z)^t$~\cite{Koba:1969rw} times the MPL generating series.
It is worth highlighting that the open- and closed-string integrals above are not the amplitudes in AdS per se, but the latter may be computed from the former via an inner-product scheme discussed in~\cite{Alday:2025cxr}.

\paragraph{Twisted de Rham theory}\!\!\!\!is another crucial element for understanding the KLT relations.
This mathematical framework~\cite{deligne1970equations,10.3792/pjaa.58.97,10.3836/tjm/1270214894,10.3836/tjm/1270214323,MR841131,MR843436,mana.19941660122,mana.19941680111} (for textbook treatments also see \cite{yoshida1997hypergeometric,Aomoto:2011ggg}) has been recognized by Mizera~\cite{Mizera:2017cqs,Mizera:2019gea} to be the natural unifying language for all of the ingredients of the KLT relations~\eqref{eq:KLT}, including the KLT kernel.
The essential idea of this formalism, which we review in some detail in \sec{sec:deRham}, is that integrals of multivalued functions, such as \eqn{eq:VenezianoAmplitude}, require a refinement of the standard de Rham theory of differential forms and chains, so as to consistently account for the branch data due to the multivaluedness.
This is accomplished by transferring the multivaluedness information from the cohomology group element (cocycle) to the homology group element (cycle) within the usual cycle–cocycle bilinear pairing.
Schematically, one replaces
\begin{equation}
\braket{[C],[{\cal I} \omega]} ~\xrightarrow[{\cal I}\text{ is multivalued}]{}~ \braket{[C\!\otimes {\cal I}],[\omega]}_\tau = \int_C {\cal I}(z)\,\omega(z) ,
\label{eq:SchematicTwist}
\end{equation}
where $C$ is a closed-contour representative of $H_m(\mathcal{M})$, ${\cal I}(z)\,\omega(z)$ is a closed-form representative of cohomology $H^m(\mathcal{M})$, and all multivaluedness is encoded in the function ${\cal I}(z)$.
The subscript $\tau$ denotes the twist, which renders the right-hand side well-defined even when the untwisted pairing is not.
This modification naturally leads to twisted homologies and cohomologies and bilinear pairings between them.

\begin{figure}[t]
\centering
\includegraphics[width=0.6\textwidth]{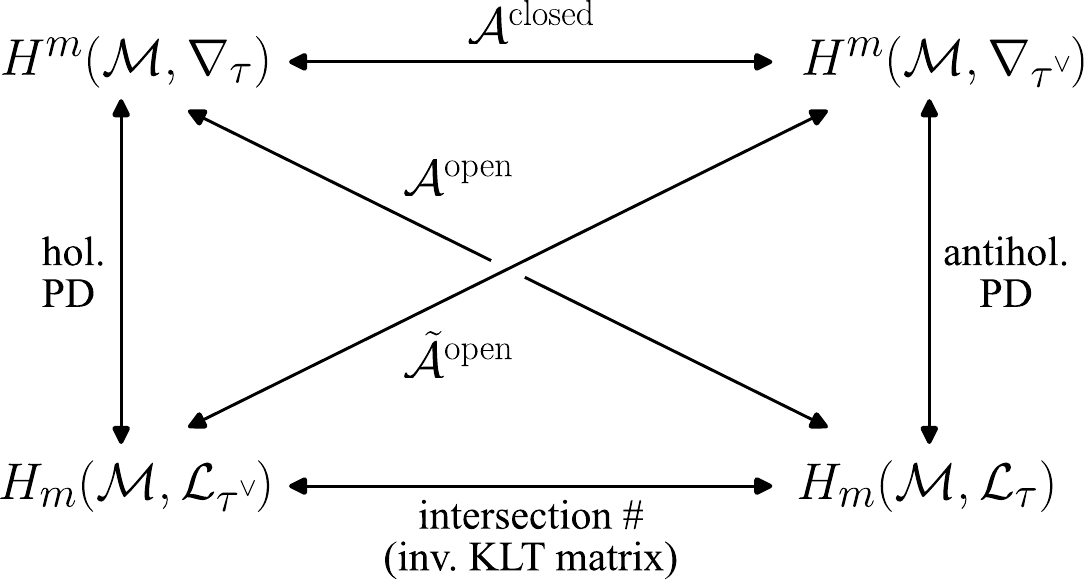}
\vspace{-5pt}
\caption{All relevant pairings of twisted de Rham theory as applied to string amplitudes.}
\label{fig:Dualities}
\end{figure}

In this formalism, tree-level open-string amplitudes are integrals over suitable twisted cycles, and the integrands are multivalued --- exclusively due to the Koba–Nielsen factor, \eg $z^s (1-z)^t$ in \eqn{eq:VenezianoAmplitude}.
Closed-string amplitudes, on the other hand, arise as canonical pairings between twisted cocycles.
Owing to Poincar\'e duality, these pairings are all interdependent, and the KLT relations follow directly from their \emph{twisted period relations}, see \fig{fig:Dualities}.
In this way, the KLT kernel is naturally interpreted as the inverse of the intersection pairing between the twisted cycles \cite{mana.19941660122,mana.19941680111,Mimachi:2002gi}.
Importantly, this avoids explicit manipulations of special functions, and the KLT kernel is instead derived from the geometric intersection points of twisted cycles.
This interpretation not only clarifies the structure of the KLT double copy but also suggests a route beyond flat space.

\paragraph{Noncommutative twisted de Rham theory}\!\!\!\!is in our opinion the natural framework for the double copy in AdS.
The aim of this work is to substantiate this claim by setting up the new noncommutative formalism and providing a first-principles derivation of the AdS-curvature-induced four-point double-copy relation~\eqref{eq:AdS_double_copy}, which is conceptually different from the original analysis of the resulting special functions~\cite{Alday:2025bjp}.

There is actually an apparent obstruction to applying twisted de Rham theory to integrals of the type~\eqref{eq:open_string_BB}.
The issue is that the twist $\tau(z)= d{\cal I}(z)/{\cal I}(z)$, which is the crucial element of this framework, must be single-valued despite ${\cal I}(z)$ being multivalued.
This works for the Koba-Nielsen factor $z^s(1-z)^t$ but does not for the MPLs~$L_w(x)$!
Fortunately, once the latter are combined into the MPL generating function~\eqref{eq:GenFunctionL}, it does work --- in the noncommutative sense! This is what we exploit in this work.
Put another way, working with MPL generating functions makes the MPL monodromies multiplicative, just as that for the Koba-Nielsen factors.
Our central claim is that the appearance of MPLs and SVMPLs, which multiply the Koba–Nielsen factor in AdS, and their relationship via curvature-dependent KLT kernels can be described geometrically using noncommutative twisted de Rham theory, which we set up in \sec{sec:AdSDC}.
The twist then incorporates both $\alpha'$-effects (from the Koba-Nielsen factor) and AdS curvature corrections.

Within this framework, AdS open-string integrals correspond to integrals of closed differential forms (twisted cocycles) over twisted cycles, while closed-string integrals arise as bilinear pairings of those open-string differential forms.
The KLT kernel is then interpreted geometrically via an intersection pairing between twisted cycles, encoding the monodromies of the underlying multivalued integrands.
These pairings produce precisely the bilinear structures relating open- and closed-string AdS building blocks.
The resulting picture hints at the AdS double-copy relations of~\cite{Alday:2025bjp} being neither accidental nor specific to particular examples, but instead reflecting a robust geometric structure underlying string perturbation theory on curved backgrounds.

%%%%%%%%%%%%%%%%%%%%%%%%%%%%%%%%%%%%%%%%%%%%%%%%%%%
\section{Twisted de Rham theory}
\label{sec:deRham}
%%%%%%%%%%%%%%%%%%%%%%%%%%%%%%%%%%%%%%%%%%%%%%%%%%%
In this section, we review the mathematical preliminaries of twisted de Rham theory.
We do this so as to reinterpret the open-string integrals like the one in \eqn{eq:open_string_AdS} as \emph{periods}
--- pairings between cohomology classes (representing integrands) and homology classes (representing integration contours).
Since our integrands are singular and multivalued, the relevant cohomologies and homologies turn out to be the ``twisted'' versions thereof~\cite{Mizera:2017cqs}.

We are interested in integrals of the type
\begin{equation}
\int_C {\cal I}(z)\,\omega(z) ,
\label{eq:TypicalIntegral}
\end{equation}
where ${\cal I}$ is a multivalued function on an $m$-dimensional complex manifold ${\cal X}$ ($2m$-real-dimensional),
$\omega \in \Omega^{m,0}({\cal X})$ is a single-valued holomorphic top form,
and $C \in C_m({\cal X},\mathbb{Z})$ is a geometrical $m$-chain (\ie a ``contour'' or a linear combination thereof with integer weights).
In the current discussion, the multivaluedness of ${\cal I}$ arises both from the non-integer powers in the Koba-Nielsen factor~\cite{Koba:1969rw} and the logarithms due to the AdS curvature corrections.
For the purposes of this section, we are going to use a simple running example:
\begin{equation}
{\cal I}(z) = z^s (1-z)^t , \qquad m=1 , \qquad {\cal X} = \mathbb{C} ,
\label{eq:FlatSpaceExample}
\end{equation}
which is relevant for the four-point open-string amplitude~\eqref{eq:VenezianoAmplitude} in flat space.
Let ${\cal D}$ denote the singular locus of~${\cal I}$, \ie its poles and branch ``points'', which in general are complex codimension-1 submanifolds of ${\cal X}$.
We remove them and define the punctured manifold ${\cal M}:={\cal X} \setminus {\cal D}$.
On this noncompact manifold, ${\cal I}$ is defined everywhere but is still multivalued.

\paragraph{Local cycle-cocycle pairing.}
In any open simply connected patch $U \subset {\cal M}$, however, we may choose a branch and treat it as a single-valued holomorphic function~${\cal I}_U(z)$.
Moreover, we can view forms~$\omega$, even with poles in ${\cal D}$, as holomorphic on all of ${\cal M}$.
Then the whole integrand constitutes a top holomorphic form on $U\subset{\cal M}$ that is automatically closed, $d({\cal I}_U \omega)=0$.
Therefore, it can be viewed as a representative of the cohomology class $[{\cal I}_U \omega] \in H^m(U)$, \ie a cocycle defined up to the addition of an exact form, ${\cal I}_U \omega \cong {\cal I}_U \omega + d\eta$.
If the integration contour~$C$ happens to be a cycle, $\partial C = 0$, that is entirely in~$U$, it represents a homology class $[C] \in H_m(U,\mathbb{Z})$.
Then the standard bilinear pairing for such classes on~$U$ is $\braket{[C],[{\cal I}_U \omega]} := \int_C {\cal I}_U \omega$.
Its consistency follows from Stokes'~theorem:
\begin{subequations}
\begin{align}
\label{eq:PDconsistencyChain}
\int_{C+\partial D}\!{\cal I}_U \omega
 = \int_C\!{\cal I}_U \omega\,+ \int_D\!d({\cal I}_U \omega) & = \braket{[C],[{\cal I}_U \omega]} \qquad
\forall D \in C_{m+1}(U,\mathbb{Z}) , \\
\label{eq:PDconsistencyForm}
\int_C ({\cal I}_U \omega+d\eta)\,
 = \int_C\!{\cal I}_U \omega\,+ \int_{\partial C}\!\eta & = \braket{[C],[{\cal I}_U \omega]} \qquad
\forall \eta \in \Omega^{m-1}(U) .
\end{align} \label{eq:PDconsistency}%
\end{subequations}
The last equality, however, can be invalidated by the noncompactness of~${\cal M}$.
Indeed, a nonzero boundary term may arise from the punctures and/or $\infty$ if $\partial C \cap \partial U \cap ({\cal D}\cup\{\infty\}) \neq 0$ --- without formally violating the closedness condition $\partial C=0$ within $U$.
So we need extra assumptions to rule out such boundary terms.
We will do this by henceforth assuming
\begin{equation}
\lim_{z\to z_0}{\cal I}(z) = 0 \quad \forall z_0\in{\cal D}\cup\{\infty\}
\label{eq:BoundaryAssumption}
\end{equation}
and incorporating the multivalued function~${\cal I}(z)$ into the notion of a chain.
In the example case~\eqref{eq:FlatSpaceExample}, this translates to ${\rm Re}(s),{\rm Re}(t)>0$ and suitable boundary conditions for~$\omega$.

%%%%%%%%%%%%%%%%%%%%%%%%%%%%%%%%%%%%%%%%%%%%%%%%%%%
\subsection{Twisted cycles and cocycles}
\label{sec:Twisted}
%%%%%%%%%%%%%%%%%%%%%%%%%%%%%%%%%%%%%%%%%%%%%%%%%%%

\paragraph{Twisted exterior derivative.}
Let us extend our discussion to more general (lower-dimensional) versions of the integral~\eqref{eq:TypicalIntegral} with a $p$-form $\omega$ and a $p$-chain~$C$ still inside~$U$.
The closedness statement for the complete integrand then amounts to the vanishing of
\begin{equation}
d({\cal I}_U\omega)
 = {\cal I}_U \big(d + d \log\,{\cal I}\wedge\;\!\!\big) \omega
 =: {\cal I}_U \nabla_\tau \omega , \qquad \text{\ie} \qquad
\nabla_\tau := d + \tau \wedge , \qquad
\tau := \frac{d{\cal I}}{{\cal I}} .
\label{eq:TwistedDifferential}
\end{equation}
Here we have defined ``connection'' $\nabla_\tau$ that involves the \emph{rank-1 twist} $\tau := d\log{\cal I}$.
For instance, the integrand~\eqref{eq:FlatSpaceExample} implies $\tau = s\,dz/z + t\,dz/(z-1)$.
Such single-valuedness of the twist is actually the standard assumption~\cite{mana.19941660122}, so $\tau$ may be extended to the entire manifold regardless of any branch choices.
Due to $d\tau=0$, the twisted differential~$\nabla_\tau$ is automatically nilpotent, $\nabla_\tau^2=0$,
and the connection is flat.
Since $\nabla_\tau \omega$ is defined globally for holomorphic forms, we may speak of \emph{twisted closedness} on ${\cal M}$, $\nabla_\tau\,\omega = 0$, which will replace the ordinary closedness condition $d\omega = 0$ in the presence of the multivalued factor~${\cal I}$.
The \emph{twisted cohomology group} is then defined in the standard way:
\begin{equation}
H^p({\cal M},\nabla_\tau) := {\rm Ker}\,\nabla_\tau /\,{\rm Im}\,\nabla_\tau\,|_{\Omega^p({\cal M})} ,
\label{eq:CohomologyGroup}
\end{equation}
\ie all $\nabla_\tau$-closed $p$-forms up to the addition of an exact form $\nabla_\tau\eta$, where $\eta\in\Omega^{p-1}({\cal M})$.
This global exterior-derivative switch $d \to \nabla_\tau$ thus puts us a step closer to pushing the multivalued factor to the homology side of the cycle-cocycle pairing, as advertised in \eqn{eq:SchematicTwist}.

\paragraph{Multivaluedness algebra.}
We still need an easy way of dealing with the multivaluedness of ${\cal I}(z)$ in a global fashion.
So instead of sticking to a single simply connected chart, we cover ${\cal M}$ by simply connected open sets $\{U_j\}$ and on each choose its own specific branch ${\cal I}_j(z)$.
Note that all branches satisfy $d{\cal I}_j=\tau {\cal I}_j$ with the same single-valued twist~$\tau$. 
This differential equation admits the formal solution $c\,{\cal I}$ with a constant $c\in\mathbb{C}$.
This solution space is one-dimensional, and on the overlap $U_j \cap U_k$ the two branches differ by a nonzero constant --- the \emph{monodromy factor}:
\begin{equation}
{\cal I}_j(z) = g_{jk}\,{\cal I}_k(z), \quad \text{where} \quad g_{jk} \in \mathbb{C}\!\setminus\!\{0\} \quad \forall z \in U_j \cap U_k .
\label{eq:MonodromyFactor}
\end{equation}
These transition-function factors satisfy the group relations
$g_{kj} = 1/g_{jk}$ and $g_{ij} g_{jk} = g_{ik}$ (no summation).
The collection of open patches and monodromies on their overlaps is known as a \emph{rank-1 local system}~${\cal L}_\tau$.
This way of encoding the branch-choice information on ${\cal M}$ is an algebraic alternative to thinking of the integration contour~$C$ as placed on the geometrically much more complicated Riemann ``surface'' that covers ${\cal M}$ (at least) as many times as there are branches --- the so-called universal cover.
For example, in the case of the function~\eqref{eq:FlatSpaceExample},
it is convenient to label nontrivially distinct patches and branches by a pair of integers and define
\begin{equation}
{\cal L}_\tau := \{U_{jk}\subset\mathbb{C}\!\setminus\!\{0,1\},~g_{(jk)(j'k')} = e^{2\pi i[(j-j')s+(k-k')t]} | j,j',k,k'\in\mathbb{Z}\} .
\label{eq:MonodromyFactorExample}
\end{equation}
A more explicit relationship between the local system and the twist is given by the map~\cite{Mizera:2019gea}
\begin{equation}
{\cal L}_\tau : C \mapsto \exp\!\int_C\!\tau , \quad \text{\eg}\!\quad
\tau = \bigg[\frac{s}{z} + \frac{t}{z-1}\bigg]dz ~~\Rightarrow~~
{\cal L}_\tau(\circlearrowleft_0) = e^{2\pi i s} ,~~
{\cal L}_\tau(\circlearrowleft_1) = e^{2\pi i t} .
\label{eq:LocalSystem}
\end{equation}
Here we have computed the monodromy factors of the above system~\eqref{eq:MonodromyFactorExample} for two basis loops around the punctures.
Note that $\tau$, if considered on the Riemann sphere $\mathbb{CP}^1 = \mathbb{C}\cup\{\infty\}$, has another pole at infinity with residue $u:=-s-t$ , for which ${\cal L}_\tau(\circlearrowleft_\infty) = e^{2\pi i u}$.
%$= 1/\big[{\cal L}_\tau(\circlearrowleft_0) {\cal L}_\tau(\circlearrowleft_1)\big]$.

\paragraph{Shifting ${\cal I}$ to chains.}
We have seen that the local system provides a way to encode the inherently geometrical information about ${\cal I}$'s branch choice.
It is then natural to think of it together with the contour~$C$ of integration and thus view the selected value of ${\cal I}(z)$ as the coefficient that multiplies the infinitesimal simplex (piece of $C$ obtained via triangulation) that passes through~$z$.
In view of this, a \emph{twisted chain} with local coefficients is defined as
\begin{equation}
C_p({\cal M},\mathbb{Z}) \ni~ C = \sum_a \sigma_a \qquad \Rightarrow \qquad
\mathscr{C} := \sum_a\,\overbrace{\!\!\!\underbrace{\sigma_a\!}_{\substack{\text{simplex}\\\text{on }{\cal M}}}\!\!\otimes\!\!\!\underbrace{{\cal I}_{\sigma_a}}_{\substack{\text{branch}\\\text{on simplex}}}\!\!\!\!\!\!}^{\text{twisted simplex}} ~\in C_p({\cal M},{\cal L}_\tau) ,
%\qquad \big(=:C\otimes{\cal I}_C\big) ,
\label{eq:TwistedChain}
\end{equation}
where each $\sigma_a$ is a $p$-simplex and ${\cal I}_{\sigma_a}$ is the branch of ${\cal I}$ chosen on it.
Whenever we wish to be more explicit, we will also slightly abuse the notation and write $C\otimes {\cal I}_C$ for twisted chains instead of~$\mathscr{C}$, with the understanding that the tensor product acts locally and depends on the local branch choice at each point of~$C$.

\paragraph{Twisted boundary operator.}
Now that we have twisted chains $C\otimes{\cal I}_C$ and the local system ${\cal L}_\tau$, which dictates how the local coefficients ${\cal I}(z)$ may change branch along the contour~$C \ni z$, we still need the corresponding boundary operator.
It is sufficient to define it on a generic $p$-simplex $\sigma=\braket{0,\!1,\!2,\!\cdots\!,p}$ spanned by $p\;\!{+}1$ nearby points:
\begin{equation}\!\!
\partial_\tau \big(\braket{0,\!1,\!2,\!\cdots\!,p} \otimes {\cal I}_{\braket{0,1,2,\cdots\;\!\!,p}} \big)
:= \sum_{j=0}^p (-1)^j \braket{0,\!1,\!2,\! \cdots\!,j\;\!{-}1,j\;\!{+}1,\!\cdots\!,p} \otimes {\cal I}_{\braket{0,1,2,\cdots\;\!\!, j-1,j+1,\cdots\;\!\!,p}} ,
\label{eq:TwistedBoundarySimplex}
\end{equation}
where the $(p\;\!{-}1)$-simplices on the right-hand side inherit the same branch of the coefficient function~${\cal I}$.
This definition mimics that of the usual boundary operator~$\partial$ and may be compressed to $\partial(\sigma\otimes{\cal I}_\sigma) = \partial\sigma \otimes {\cal I}_{\partial\sigma}$.
So the nilpotence $\partial_\tau^2 = 0$ follows directly from $\partial^2=0$.
A new feature, however, is that a boundary may now become zero due to the vanishing of~${\cal I}(z)$.
For instance, if we take $\braket{0,1}$ to be the 1-simplex from $z=0$ to $z=1$ and dress it with the integrand~\eqref{eq:FlatSpaceExample}, we find that the latter makes it a \emph{twisted cycle}:
\begin{equation}
\partial\,\braket{0,1} = \braket{1} - \braket{0} \qquad \text{vs.} \qquad
\partial_\tau \big(\braket{0,1} \otimes z^s (1{-}z)^t\big) = 0 .
\label{eq:FlatSpaceExampleCycle}
\end{equation}
More generally, this feature works very well together with our boundary-behavior assumption~\eqref{eq:BoundaryAssumption}.
Indeed, integration contours~$C$ with endpoints in the divisor~${\cal D}$ (or at $\infty$) are now guaranteed to be closed on par with genuine loops.
As we saw in \eqn{eq:PDconsistencyForm}, boundary terms due to such contours could interfere with Poincar\'e duality even despite their endpoints being punctured out of our manifold ${\cal M}$.
The advantage of the twisted boundary operator~\eqref{eq:TwistedBoundarySimplex} is that it sets such problematic boundaries to zero purely algebraically (assuming tame behavior of the single-valued integrand~$\omega$ at all punctures and~$\infty$).

Any twisted cycle defined in this way, $\partial_\tau (C\otimes{I}_C) = 0$, constitutes a representative of the \emph{twisted homology group}, which is defined as
\begin{equation}
H_p({\cal M},{\cal L}_\tau) := {\rm Ker}\,\partial_\tau/\,{\rm Im}\,\partial_\tau\,|_{C_p({\cal M},{\cal L}_\tau)} ,
\label{eq:HomologyGroup}
\end{equation}
\ie all $\partial_\tau$-closed $p$-chains $\mathscr{C}$ up to the addition of an exact chain $\partial_\tau \mathscr{O}$, where $\mathscr{O}$ is a $(p\;\!{+}1)$-chain.
Such homologies are qualitatively different from their non-twisted counterparts.
For instance, the more standard homology of the twice punctured complex plane, relevant for contour integrals of single-valued forms, such as $s\,dz/z+t\,dz/(z-1)$ with poles at $\{0,1,\infty\}$, involves the field of integers, which count anticlockwise loops around the poles:
\begin{equation}
H_1(\mathbb{C}\!\setminus\!\{0,1\},\mathbb{Z})
 = \{ n_0\text{ cycles }\!\circlearrowleft_0\,\oplus\,n_1\text{ cycles }\!\circlearrowleft_1 | n_0,n_1\in\mathbb{Z} \}
 \cong \mathbb{Z}^2 .
\label{eq:FlatSpaceHomology}
\end{equation}
Now in the twisted homology, such loops cease being cycles altogether, as their (arbitrarily chosen) endpoints~$z$ contribute different weights to their boundaries:
\begin{equation}
\partial_\tau \big(\!\circlearrowleft_0\otimes\,{\cal I}_{\circlearrowleft_0}\big) = z \otimes [e^{2\pi i s}\!-1]{\cal I}(z) , \qquad \quad
\partial_\tau \big(\!\circlearrowleft_1\otimes\,{\cal I}_{\circlearrowleft_1}\big) = z \otimes [e^{2\pi i t}\!-1]{\cal I}(z) .
\label{eq:TwistedNonCycles}
\end{equation}
These contours now make up the local system~\eqref{eq:LocalSystem},
whereas the homology consists of new twisted cycles between the punctures ${\cal D}\cup\{\infty\}$,
such as the one in \eqn{eq:FlatSpaceExampleCycle}.
Moreover, due to the inclusion of local weights the number field for twisted homologies is necessarily~$\mathbb{C}$.
For the example~\eqref{eq:FlatSpaceExample}, it can be shown (see \eg \cite{Mizera:2019gea}) that three basic intervals $(0,1)$, $(1,\infty)$ and $(\infty,0)$ are linearly dependent as twisted cycles, and the twisted homology is simply
\begin{equation}
H_1(\mathbb{C}\!\setminus\!\{0,1\},{\cal L}_\tau)
 = \big\{ (0,1) \otimes cz^s (1-z)^t | c\in\mathbb{C} \big\}
 \cong \mathbb{C} .
\label{eq:FlatSpaceHomologyTwisted}
\end{equation}

%%%%%%%%%%%%%%%%%%%%%%%%%%%%%%%%%%%%%%%%%%%%%%%%%%%
\subsection{Network of dualities}
\label{sec:Dualities}
%%%%%%%%%%%%%%%%%%%%%%%%%%%%%%%%%%%%%%%%%%%%%%%%%%%

We are now ready to gradually introduce all the dualities that were advertised in \fig{fig:Dualities}.
\paragraph{Twisted periods.}
The new differential and boundary operator obey their own Stokes theorem:
\begin{equation}
\int_{C\otimes{\cal I}_C}\!\!\nabla_\tau\omega\,
 = \int_{\partial_\tau(C\otimes{\cal I}_C)}\!\omega ,
\qquad
\begin{aligned}
(C\otimes{\cal I}_C) & \in C_{p+1}({\cal M},{\cal L}_\tau) , \\
\omega & \in \Omega^p({\cal M}) .
\end{aligned}
\label{eq:TwistedStokes}
\end{equation}
It can be proven using the conventional Stokes theorem for each constituent simplex:
\begin{equation}
\int_\mathscr{C}\!\nabla_\tau\omega
 = \sum_a\!\int_{\sigma_a}\!\!\!{\cal I}_{\sigma_a}\!\nabla_\tau\omega 
 = \sum_a\!\int_{\sigma_a}\!\!\!d\big({\cal I}_{\sigma_a} \omega\big)
 = \sum_a\!\int_{\partial\sigma_a}\!\!\!{\cal I}_{\sigma_a} \omega
 = \sum_a\!\int_{\partial\sigma_a}\!\!\!{\cal I}_{\partial\sigma_a} \omega
 =\!\int_{\partial_\tau\mathscr{C}}\!\omega .
\label{eq:TwistedStokesProof}
\end{equation}
Equipped with these mathematical structures, we can finally introduce the global bilinear period pairing between the twisted homology and cohomology groups:
\begin{equation}
H_p({\cal M},{\cal L}_\tau) \times H^p({\cal M},\nabla_\tau) \to \mathbb{C} : \qquad
\braket{[\mathscr{C}], [\omega]}_\tau := \int_{C\otimes{\cal I}_C}\!\!\omega ,
\label{eq:Period}
\end{equation}
where $\mathscr{C}=C\otimes{\cal I}_C$.
Its consistency follows from the above version of Stokes' theorem:
\begin{subequations}
\begin{align}
\label{eq:PDconsistencyChainTwisted}
\int_{\mathscr{C}+\partial_\tau\mathscr{D}}\!\omega
 = \int_\mathscr{C}\!\omega\,+ \int_\mathscr{D}\!\nabla_\tau \omega & = \braket{[\mathscr{C}],[\omega]}_\tau \qquad
\forall \mathscr{D} \in C_{p+1}({\cal M},{\cal L}_\tau) , \\
\label{eq:PDconsistencyFormTwisted}
\int_\mathscr{C}(\omega+\nabla_\tau\eta)\,
 = \int_\mathscr{C}\!\omega\,+ \int_{\partial_\tau\mathscr{C}}\!\eta & = \braket{[\mathscr{C}],[\omega]}_\tau~\qquad
\forall \eta \in \Omega^{p-1}({\cal M}) .
\end{align} \label{eq:PDconsistencyTwisted}%
\end{subequations}

For an example of a period~\eqref{eq:Period}, let us choose the twisted cycle from the 1-complex-dimensional homology~\eqref{eq:FlatSpaceHomology} and a $\nabla_\tau$-closed form $\omega = dz/z + dz/(1-z) = d\log[z/(1-z)]$.
Note that the latter is $d$-exact but not $\nabla_\tau$-exact, otherwise the following would be zero:\footnote{Starting from \eqn{eq:VenezianoAmplitudePeriod} and henceforth, we drop the square brackets in all pairings, as soon as their invariance with respect to the choice of the (co)homology class representative is established.
}
\begin{equation}
\Big\langle \braket{0,1} \otimes z^s (1-z)^t, d\log\frac{z}{1-z} \Big\rangle_\tau
 = \int_0^1\!z^{s-1} (1-z)^{t-1} dz = B(s,t) = \frac{\Gamma(s)\Gamma(t)}{\Gamma(s+t)} .
\label{eq:VenezianoAmplitudePeriod}
\end{equation}
This result is actually (proportional to) the color-ordered version of Veneziano's open-string amplitude~\cite{Veneziano:1968yb} discussed around \eqn{eq:VenezianoAmplitude}.

Since both $H_p({\cal M},{\cal L}_\tau)$ and $H^p({\cal M},\nabla_\tau)$ are finite-dimensional and are dual to each other, their dimensions coincide.
Their relationship can be captured by a period matrix~$P$, defined for a basis~$\{\mathscr{E}^j=E^j\otimes{\cal I}_{E^j}\}$ for twisted cycles and $\{\varepsilon_j\}$ for twisted cocycles:
\begin{equation}
P_k{}^j := \braket{\mathscr{E}^j,\varepsilon_k}_\tau
 = \int_{E^j\otimes{\cal I}_{E^j}}\!\!\varepsilon_k
\qquad \Rightarrow \qquad
\braket{\mathscr{C},\omega}_\tau = [\omega]^k P_k{}^j [\mathscr{C}]_j .
\label{eq:PeriodMatrix}
\end{equation}
Namely, this matrix records all values of the integrals of the chosen cohomology basis over the chosen homology basis.
Note the picked order of indices, which will be more important in \sec{sec:AdSDC}.
We choose to think of basis changes as multiplication of~$P$ by the corresponding basis-change matrices on the right (homology basis) or left (cohomology basis):
\begin{equation}
\mathscr{E}'^j = \mathscr{E}^k A_k{}^j , \qquad
\varepsilon'_j = B_j{}^k \varepsilon_k 
\qquad \Rightarrow \qquad
P'{}_l{}^i = B_l{}^k P_k{}^j  A_j{}^i .
\label{eq:PeriodMatrixTransform}
\end{equation}

\paragraph{Dual periods.}
For simplicity, we refocus on maximal holomorphic forms $\omega \in \Omega^{m,0}({\cal M})$, the related cohomology $H^m({\cal M},\nabla_\tau)$ and its dual homology $H_m({\cal M},{\cal L}_\tau)$.
It is natural to also include their antiholomorphic versions $H_m({\cal M},{\cal L}_{\bar\tau}) \cong H^m({\cal M},\nabla_{\bar\tau})$ into the picture, for which a similar construction must hold.
In fact, this antiholomorphic construction does not have to involve the identical twist $\bar{\tau} = d\log{\cal I}(\bar{z})$.
Instead, it may start with another multivalued function, which we will denote as~${\cal I}^\vee(\bar{z})$, and use the twist $\tau^\vee := d\log{\cal I}^\vee(\bar{z})$, as long as its punctures are the same and give the same manifold~${\cal M}$.
So let us assume the whole preceding construction holds for the antiholomorphic period pairing
\begin{equation}
H_m({\cal M},{\cal L}_{\tau^\vee}\!) \times H^m({\cal M},\nabla_{\tau^\vee}\!) \to \mathbb{C} : \qquad
\braket{[\mathscr{C}], [\psi]}_{\tau^\vee} := \int_{C\otimes{\cal I}_C^\vee}\!\!\psi(\bar{z}) .
\label{eq:PeriodDual}
\end{equation}
Choosing a basis $\{\tilde\varepsilon^j\}$ for antiholomorphic twisted cocycles and allowing for a new basis~$\{\tilde{\mathscr{E}}_j=\tilde{E}_j\otimes{\cal I}_{\tilde{E}_j}^\vee\}$ for twisted cycles,
we define the dual period matrix $\widetilde{P}$:
\begin{equation}
\widetilde{P}_j{}^k := \braket{\tilde{\mathscr{E}}_j,\tilde\varepsilon^k}_{\tau^\vee}
 = \int_{\tilde{E}_j\otimes{\cal I}_{\tilde{E}_j}^\vee}\!\!\tilde\varepsilon^k(\bar{z})
\qquad \Rightarrow \qquad
\braket{\mathscr{C},\psi}_{\tau^\vee}
 = [\mathscr{C}]^j \widetilde{P}_j{}^k [\psi]_k .
\label{eq:PeriodMatrixDual}
\end{equation}

\paragraph{Poincar\'e duality.}
On orientable manifolds, we may complement these two dualities by the following bilinear pairing of the holomorphic and antiholomorphic twisted cocycles:
\begin{equation}
H^m({\cal M},\nabla_\tau) \times H^m({\cal M},\nabla_{\tau^\vee}\!) \to \mathbb{C} : \qquad
\bbraket{[\omega],[\psi]}_{\cal M} := \int_{\cal M}\!{\cal I}^\vee(\bar{z})\,{\cal I}(z)\,\omega(z)\wedge\psi(\bar{z}) .
\label{eq:CohomologyPairing}
\end{equation}
Its invariance with respect to the representative change follows from Stokes' theorem~\eqref{eq:TwistedStokes}:
\begin{align}
\label{eq:CohomologyPairingProof}
\bbraket{[\omega+\!\nabla_\tau\eta],[\psi]}_{\cal M} - \bbraket{[\omega],[\psi]}_{\cal M} &
 = \int_{\cal M}\!{\cal I}^\vee {\cal I}\,(\nabla_\tau\eta)\wedge\psi
 = \int_{\cal M}\!{\cal I}^\vee d({\cal I}\,\eta)\wedge\psi \\
 = (-1)^{m+1}\!\!\int_{\cal M}\! {\cal I}\,\omega \wedge d({\cal I}^\vee\psi) &
 = (-1)^{m+1}\!\!\int_{\cal M}\! {\cal I}^\vee {\cal I}\,\omega \wedge \nabla_{\tau^\vee}\psi = 0 , \qquad
\forall \eta \in \Omega^{m-1}({\cal M}) . \nn
\end{align}
Hence this cocycle pairing establishes the duality between the holomorphic and antiholomorphic cohomologies.
An example of such a pairing for ${\cal I}^\vee(\bar{z})=\overline{{\cal I}(z)}=\bar{z}^s(1-\bar{z})^t$ is
\begin{equation}\!\!\!
\Big\langle\!\!\Big\langle d\log\frac{z}{1{-}z}, d\log\frac{\bar{z}}{1{-}\bar{z}} \Big\rangle\!\!\Big\rangle_{\!\cal M}\!
 =\!\int_{\mathbb{C}\setminus\{0,1\}}\!\!\!\!\!\!\!\!\!dz \wedge d\bar{z} |z|^{2s-2} |1{-}z|^{2t-2}
 = -\frac{2\pi i\,\Gamma(s)\Gamma(t)\Gamma(1{-}s{-}t)}{\Gamma(s+t)\Gamma(1{-}s)\Gamma(1{-}t)} ,
\label{eq:VirasoroShapiroAmplitudePeriod}
\end{equation}
which is the Virasoro-Shapiro closed-string amplitude~\eqref{eq:VirasoroShapiroAmplitude} for four tachyons~\cite{Virasoro:1969me,Shapiro:1969km}.
The cohomological pairing can be encapsulated in the cocycle matrix $F$:
\begin{equation}\!
F_j{}^k := \bbraket{\varepsilon_j,\tilde\varepsilon^k}_{\cal M}
 = \int_{\cal M}\!{\cal I}^\vee(\bar{z})\,{\cal I}(z)\,\varepsilon_j(z)\wedge\tilde\varepsilon^k(\bar{z})
\quad\,\Rightarrow\,\quad
\bbraket{[\omega],[\psi]}_{\cal M} := [\omega]^j F_j{}^k [\psi]_k .
\label{eq:CocycleMatrix}
\end{equation}

\begin{subequations}
The above three pairings result in Poincar\'e duality, which provides two isomorphisms: one between holomorphic cocycles in $H^m({\cal M},\nabla_\tau)$ and dual cycles in $H_m({\cal M},{\cal L}_{\tau^\vee}\!)$:
\begin{equation}\!\!\!
H^m({\cal M},\nabla_\tau) \leftrightarrow H_m({\cal M},{\cal L}_{\tau^\vee}\!)\!: \quad
\braket{\tilde{\mathscr{C}}(\omega),\psi}_{\tau^\vee} = \bbraket{\omega(\tilde{\mathscr{C}}),\psi}_{\cal M} \quad
\forall [\psi]\!\in\!H^m({\cal M},\nabla_{\tau^\vee}\!) ,\!
\label{eq:PoincareDualityHol}
\end{equation}
and another between antiholomorphic cocycles in $H^m({\cal M},\nabla_{\tau^\vee}\!)$ and cycles in $H_m({\cal M},{\cal L}_\tau)$:
\begin{equation}
H^m({\cal M},\nabla_{\tau^\vee}\!) \leftrightarrow H_m({\cal M},{\cal L}_\tau)\!: \quad
\braket{\mathscr{C}(\psi),\omega}_\tau = \bbraket{\omega,\psi(\mathscr{C})}_{\cal M} \quad
\forall [\omega]\!\in\!H^m({\cal M},\nabla_\tau) .
\label{eq:PoincareDualityAntihol}
\end{equation} \label{eq:PoincareDuality}%
\end{subequations}
These isomorphisms are readily expressed in terms of the cocycle and period matrices:
\begin{equation}
[\tilde{\mathscr{C}}]^j(\omega) = [\omega]^l\,F_l{}^k\,\widetilde{P}^{-1}{}_k{}^j , \qquad \quad
[\mathscr{C}]_j(\psi) = P^{-1}{}_j{}^k\,F_k{}^l\,[\psi]_l .
\label{eq:PoincareDualityMatrices}
\end{equation}

\paragraph{Intersection numbers.}
The network of pairings described above is still lacking one between the two homologies $H_m({\cal M},{\cal L}_{\tau^\vee}\!)$ and $H_m({\cal M},{\cal L}_\tau)$.
In fact, it is already implied by the other dualities, so we may use the following ``intersection-number'' matrix
\begin{align}
I_j{}^k & := \braket{\tilde{\mathscr{E}}_j,\mathscr{E}^k}_{\cal M}
:= \bbraket{\omega(\tilde{\mathscr{E}}_j),\psi(\mathscr{E}^k)}_{\cal M}
 = \widetilde{P}_j{}^i F^{-1}{}_i{}^l P_l{}^k
\label{eq:IntersectionNumberMatrix}
\end{align}
as the definition via the basis cycles, which is then trivial to extend to other cycles by linearity
$\braket{[\tilde{{\mathscr{C}}}],[\mathscr{C}]}_{\cal M} := [\tilde{{\mathscr{C}}}]^j I_j{}^k [\mathscr{C}]_k$.
For ordinary homologies, the intersection pairing $H_m({\cal M}) \times H_m({\cal M}) \to \mathbb{Z}$ is purely geometric and corresponds to counting the intersections between closed contours with orientation-dependent signs, hence the name (see \eg \cite{Mizera:2019gea}).
For twisted homologies, the definition~\eqref{eq:IntersectionNumberMatrix} via the cohomology pairing~\eqref{eq:CohomologyPairing} may similarly be proven to be equivalent to a definition as a sum over the intersection points:
\begin{equation}\!\!
H_m({\cal M},{\cal L}_{\tau^\vee}\!) \times H_m({\cal M},{\cal L}_\tau) \to \mathbb{C} : \qquad
\braket{[\tilde{C}\otimes{\cal I}^\vee_{\tilde C}], [C\otimes{\cal I}_C]}_{\cal M} :=\!\sum_{z \in \tilde{C} \cap C}\!\!\pm \frac{{\cal I}_C(z)\,{\cal I}^\vee_{\tilde C}(z)}{|{\cal I}(z)\,{\cal I}^\vee(z)|} ,
\label{eq:IntersectionNumber}
\end{equation}
where each point is now weighted not just by a sign but also by monodromy factors.
This point is subtle, and we defer discussing it in more detail to \sec{sec:IntersectionPairing}.
So let us conclude this review by summarizing the matrix relationships between all the relevant pairings:
\begin{equation}
\begin{aligned}
I & = \widetilde{P} F^{-1} P , \\
F & = P I^{-1} \widetilde{P} ,
\end{aligned} \qquad \quad
\begin{aligned} \relax
[{\rm PD}(\tilde{\mathscr{E}}_j)](z) & = \widetilde{P}_j{}^k F^{-1}{}_k{}^l [\varepsilon_l](z) , \\
[{\rm PD}(\mathscr{E}^j)](\bar{z}) & = [\tilde\varepsilon^l](\bar{z})\,F^{-1}{}_l{}^k P_k{}^j .
\end{aligned}
\label{eq:PeriodRelations}
\end{equation}
The left-hand identities are called \emph{twisted period relations},
and on the right-hand side we have provided explicit formulae for the Poincar\'e dual cocycle class for a basis cycle.

%%%%%%%%%%%%%%%%%%%%%%%%%%%%%%%%%%%%%%%%%%%%%%%%%%%
\section{AdS double copy}
\label{sec:AdSDC}
%%%%%%%%%%%%%%%%%%%%%%%%%%%%%%%%%%%%%%%%%%%%%%%%%%%

The twisted de Rham framework has so far been set up around holomorphic differential forms $\omega(z)$ and a single scalar multivalued factor~${\cal I}(z)$.
This was found~\cite{Mizera:2017cqs} to be a natural language for describing open- and closed-string amplitudes in flat space, as well as the KLT relations~\cite{Kawai:1985xq} connecting them.
In this section, we extend this twisted de Rham picture to AdS background by proving the recently proposed AdS double copy of \cite{Alday:2025bjp}.

The immediate challenges to this originate from the fact that the AdS curvature corrections modify the multivalued function ${\cal I}(z)$ by including an infinite family of multiple polylogarithms.
Fortunately, their branch points coincide with those of the Koba-Nielsen factor.
Moreover, they can all be collected into a single MPL generating function $L(e_\ell;z)$, where $e_\ell$ are the auxiliary variables indexed by letters~$\ell$ in a finite alphabet.
At four points, they belong to the alphabet of only two letters: 0 and 1, corresponding to the branch points.
Importantly, these generators do not commute: $e_0 e_1 \neq e_1 e_0$.
This raises a string of natural questions.
What algebraic object collects these $e_\ell$'s?
How do we define the twist with non-commuting objects?
What is a twisted cycle if the new multivalued integrand ${\cal I}(z;e_\ell)$ is noncommutative?
If the order of multiplication matters, how do we pair twisted cycles and cocycles to get numbers? 

These questions naturally lead to the need for a \emph{noncommutative} ring ${\cal R}$ (the algebra of words) and modules over it (see \eg \cite{Vanhove:2018elu}).
This is because the natural algebraic environment that contains all possible ordered products of the letters, $e_0$ and $e_1$, is the free associative algebra $\mathbb{C}\langle e_0,e_1 \rangle$.
Free means that there are no relations among the letters beyond associativity: multiplication is simply concatenation of words.
This algebra consists of all finite linear combinations of words such as
\begin{equation}\!\!\!
e_\text{empty}\!\equiv\!1,~e_0,~e_1,~e_0 e_0,~e_0 e_1,~e_1 e_0,~e_1 e_1,~e_0 e_0 e_0,~\text{\etc}
\end{equation}
However, we encounter the generating function of MPLs, which is a formal infinite sum.
So finite linear combinations do not suffice, and we must pass to a completion of the free associative algebra that allows infinite sums of words.
This gives the completed free associative algebra ${\cal R}= \widehat{\mathbb{C}\langle e_0,e_1 \rangle}$, which is the algebra of noncommutative formal power series in $e_0$ and~$e_1$.

Below, we gradually introduce all the necessary ingredients for twisted de Rham theory in this noncommutative setting and use it to prove the four-point AdS double copy of \cite{Alday:2025bjp}.

%%%%%%%%%%%%%%%%%%%%%%%%%%%%%%%%%%%%%%%%%%%%%%%%%%%
\subsection{Twist and local system}
\label{sec:twist}
%%%%%%%%%%%%%%%%%%%%%%%%%%%%%%%%%%%%%%%%%%%%%%%%%%%

We consider a generic integrand of the type
\begin{equation}
\omega(z;e_\ell)\,{\cal I}(z;e_\ell) ,
\label{eq:Int_WI}
\end{equation}
which we would like to result in a period pairing in the twisted de Rham language.
The precise form of ${\cal I}(z;e_\ell)$ for the four-point AdS open-string building blocks reads
\begin{equation}
{\cal I}(z;e_0,e_1) = \underbrace{z^s(1-z)^t}_{\substack{\text{Koba-Nielsen}\\\text{factor}}} \underbrace{L(e_0,e_1;z)}_{\substack{\text{MPL}\\\text{gen. function}}} .
\label{eq:AdS_open_building_block}
\end{equation}
However, the exact form of ${\cal I}(z;e_\ell)$ will only be used in \sec{sec:DC4pt}, and until then it may be kept as an unspecified multivalued function.
Similarly, $\omega(z;e_\ell)$ in \eqn{eq:Int_WI} will be an unspecified holomorphic $m$-form of the noncommutative generators~$e_\ell$.
Since both of the factors in the integrand involve non-commuting generators, the ordering matters, and in \eqn{eq:Int_WI} we chose ${\cal I}$ to follow $\omega$.
This choice will become clear below.

As in \sec{sec:deRham}, we now consider an $m$-dimensional manifold~${\cal M}$ with punctures that correspond to the singular locus of the multivalued function~${\cal I}$.
Since all the branch points of both the Koba–Nielsen factor and the MPL generating series coincide, the singular locus in the present scenario is unchanged with respect to the flat-space case.
So at four points we have ${\cal M} = \mathbb{C}\!\setminus\!\{0,1\}$, and its geometry is unaltered from our previous running example~\eqref{eq:FlatSpaceExample}.
What changes, however, is that the multivalued function now takes values in a noncommutative ring~${\cal R}$.

\paragraph{Twisted connection.}
To define the twist, it is convenient to work with a single simply connected open patch $U\subset{\cal M}$ with a specific branch of the multivalued integrand and consider its integration over a topological $m$-cycle $C \in U$.
For a generic holomorphic $p$-form $\omega(z;e_\ell)$, the usual exterior derivative gives
\begin{equation}
d\big(\omega(z;e_\ell)\,{\cal I}(z;e_\ell)\big)
 = \big[d\omega(z;e_\ell) + (-1)^p \omega(z;e_\ell) \wedge d{\cal I}(z;e_\ell)\,{\cal I}^{-1}(z;e_\ell) \big]\,{\cal I}(z;e_\ell) .
\label{eq:sttwist}
\end{equation}
Since the ordering matters, the only constraint on ${\cal I}(z;e_\ell)$ so far is that it obeys a differential equation of the type
\begin{equation}
d{\cal I}(z;e_\ell)=\tau(z;e_\ell)\,{\cal I}(z;e_\ell) .
\label{eq:NCdiff}
\end{equation}
In other words, we define the twist as the noncommutative ``dlog'' form $d{\cal I} \times {\cal I}^{-1}$.\footnote{Here and below, the inverse in~${\cal R}$ is understood in the sense of formal series expansions, such as $\big(c + {\cal O}(e_0, e_1)\big)^{-1}=c-{\cal O}(e_0, e_1)+\big({\cal O}(e_0, e_1)\big)^2+\dots$ for some $c\in\mathbb{C}$, see \eg \app{app:Equivalence}.
}
This requirement is motivated by the four-point AdS integrand~\eqref{eq:AdS_open_building_block}.
Indeed, both the Koba-Nielsen factor and the MPL generating function satisfy this constraint individually:
\begin{subequations}
\begin{align}
d\big(z^s(1-z)^t\big) &
 = \bigg[\frac{s}{z} + \frac{t}{z-1}\bigg] dz\,z^s(1-z)^t , \\
dL(e_0,e_1;z) &
 = \bigg[\frac{e_0}{z}+\frac{e_1}{z-1}\bigg] dz\,L(e_0,e_1;z) ,
\end{align}
\end{subequations}
where the latter is the Knizhnik-Zamolodchikov equation~\cite{Knizhnik:1984nr}.\footnote{The generating series for the SVMPL ${\cal L}(e_\ell; z)$ also satisfies its own Knizhnik-Zamolodchikov equation: ${\partial}{\cal L}(e_0,e_1;z)/{\partial z}=[{e_0}\,dz/{z}+{e_1}\,dz/(z-1)]{\cal L}(e_0,e_1;z).$}
In view of \eqns{eq:sttwist}{eq:NCdiff}, we define the twisted connection
\begin{equation}
\nabla_\tau\,\omega(z;e_\ell) := d\omega(z;e_\ell) + (-1)^p\,\omega(z;e_\ell) \wedge \tau(z;e_\ell) \,.
\label{eq:TwistedDifferentialNC1}
\end{equation}
The twist $\tau(z;e_\ell)$ in the four-point case~\eqref{eq:AdS_open_building_block} reads
\begin{equation}
\tau(z;e_\ell) = \bigg[\frac{s+e_0}{z}+\frac{t+ e_1}{z-1}\bigg] dz .
\label{eq:Non_commutative_Twist}
\end{equation}
It is clearly a single-valued, albeit noncommutative, function.
So we can consider it as another constraint on ${\cal I}(z;e_\ell)$ that its dlog form is single-valued.
This allows us to extend the above definition from~$U$ to the rest of ${\cal M}$.
The twisted connection is automatically flat:
\begin{equation}
\begin{aligned}
\nabla_\tau^2 \omega & = \nabla_\tau(\nabla_\tau \omega)
 = \nabla_\tau d\omega + (-1)^p \nabla_\tau (\omega \wedge \tau) \\ &
 = (-1)^p \big[{-}d\omega \wedge \tau + d(\omega \wedge \tau) + (-1)^{p+1} \omega \wedge \tau \wedge \tau\big] = (-1)^{2p} \omega \wedge d\tau = 0 ,
\end{aligned}
\end{equation}
where in the end we have used that $\tau(z;e_\ell)$ is $d$-closed.

The requirement~\eqref{eq:NCdiff} fixes the order of the multivalued function and the form in \eqn{eq:Int_WI}.
To see this, let us consider the opposite ordering and compute
\begin{equation}
\begin{aligned}
d\big({\cal I}(z;e_\ell)\,\omega(z;e_\ell)\big) &
 = d {\cal I}(z;e_\ell) \wedge \omega(z;e_\ell) + {\cal I}(z;e_\ell) d\omega(z;e_\ell) \\ &
 =: \tau(z;e_\ell)\,{\cal I}(z;e_\ell) \wedge \omega(z;e_\ell) +  {\cal I}(z;e_\ell)\,d\omega(z;e_\ell) .
\label{eq:sttwist_Reverse}
\end{aligned}
\end{equation}
Since we do not require ${\cal I}(z;e_\ell)$ to commute with either $\tau(z;e_\ell)$ or $\omega(z;e_\ell)$, we see no consistent way of factoring out ${\cal I}(z;e_\ell)$ and defining a twisted connection.

\paragraph{Local system.}
Now let us deal with the multivaluedness of ${\cal I}(z;e_\ell)$.
We define the local system ${\cal L}_{\tau}$, which comprises the set of patches $\{U_j\}$ covering ${\cal M}$, the set of the single-valued solutions $\{{\cal I}_j(z)\}$ of the differential equation~\eqref{eq:NCdiff} on each patch $U_j$, as well as the set of transition functions $g_{jk}$ between them.
In this way, we take the twist~$\tau$ to define different branches of ${\cal I}$ as different solutions of $d{\cal I} = \tau {\cal I}$.
Since $\tau$ is on the left in this equation, the formal solution has the form
\begin{equation}
{\cal I}(z;e_\ell)\,c(e_\ell)
\end{equation}
with the ring-valued constant on the right.
Indeed, it is straightforward to check that if ${\cal I}(z;e_\ell)$ is a solution of \eqn{eq:NCdiff} then ${\cal I}(z;e_\ell)\,c(e_\ell)$ is as well:
\begin{equation}
d{\cal I}\,{\cal I}^{-1} = \tau \qquad \Rightarrow \qquad d{\cal I}\,c\,({\cal I}\,c)^{-1} = d{\cal I}\,{\cal I}^{-1} = \tau ,
\end{equation}
whereas $c(e_\ell)\,{\cal I}(z;e_\ell)$ is not unless $c(e_\ell) \in \mathbb{C}$:
\begin{equation}
d{\cal I}\,{\cal I}^{-1} = \tau \qquad \Rightarrow \qquad c\,d{\cal I} (c\,{\cal I})^{-1} = c\,\tau\,c^{-1} .
\end{equation}
In other words, our multivalued functions form a right ${\cal R}$-module.
On a given overlap~$U_j \cap U_k$, the local solutions may pick different branches of ${\cal I}$ and are related by the right-multiplicative transition functions, \ie monodromy factors:
\begin{equation}
{\cal I}_j(z;e_\ell) =  {\cal I}_k(z;e_\ell)\,g_{jk}(e_\ell) , \quad \text{where} \quad g_{jk}(e_\ell) \in {\cal R} \quad \forall z \in U_j \cap U_k .
\label{eq:MonodromyFactor2}
\end{equation}
They satisfy the group relations $g_{jk} g_{ij} = g_{ik}$ (no summation).

The right multiplication choice is in agreement with the type of multivalued functions that we are interested in.
Indeed, the monodromies of the MPL generating function around~0 and~1 are respectively \cite{MINH2000217,Brown:2004ugm}
\begin{subequations}
\begin{align}
L(e_0,e_1;z) M_0 & = L(e_0,e_1;z)\,e^{2\pi ie_0} , \\
L(e_0,e_1;z) M_1 & = L(e_0,e_1;z)\,Z(e_1,e_0)\,e^{2\pi ie_1} Z(e_0,e_1) .
\end{align} \label{eq:Monodromies01}%
\end{subequations}
Here $Z(e_0,e_1)$ is the Drinfeld associator
\begin{equation}
Z(e_0,e_1) := L(e_0,e_1;1) = 1 - \zeta(2) [e_0 e_1 - e_1 e_0] + \dots ,
\label{eq:DrinfeldAssociator}
\end{equation}
whereas $Z(e_1,e_0)$ is the inverse associator, \ie $Z(e_0,e_1)Z(e_1,e_0)=1$.
In view of noncommutativity, it is worth checking that the above transition function does not introduce nontrivial transformation rules for either the twist or the connection.
So we differentiate the relation ${\cal I}_j = {\cal I}_k g_{jk}$ and obtain
\begin{equation}\!
\tau_j:= d{\cal I}_j\,{\cal I}_j^{-1} = d{\cal I}_k\,g_{jk}\,{\cal I}_j^{-1}
    =d{\cal I}_k\,g_{jk}\,({\cal I}_k g_{jk})^{-1} = d{\cal I}_k\,g_{jk}\,g_{jk}^{-1}\,{\cal I}_k^{-1} = d{\cal I}_k\,{\cal I}_k^{-1} =:\tau_k .
\end{equation}
We have thus verified that $\tau$ is single-valued.
Therefore, the twisted connection $\nabla_\tau$ is defined globally on ${\cal M}$, and the notion of $\nabla_\tau$-closedness is independent of the chart.

%%%%%%%%%%%%%%%%%%%%%%%%%%%%%%%%%%%%%%%%%%%%%%%%%%%
\subsection{Dual twist and local system}
\label{sec:dual}
%%%%%%%%%%%%%%%%%%%%%%%%%%%%%%%%%%%%%%%%%%%%%%%%%%%
In the mathematical literature, to the best of our knowledge, the entire machinery of twisted de Rham theory is defined for a local system and its dual where the latter is obtained either through inversion ${\cal I}(z) \to 1/{\cal I}(z)$ or complex conjugation ${\cal I}(z) \to {\cal I}(\bar{z})$.
In \sec{sec:Dualities}, we saw that the latter was the natural duality in the context of string amplitudes in flat space.
In the current scenario, however, neither of these suffice, and the dual system that we will use to prove the stringy double copy with AdS curvature corrections is obtained in three steps when going from ${\cal I}$ to ${\cal I}^\vee$:
\begin{itemize}
\item complex conjugation $z \to \bar{z}$,
\item reversal of the order of letters inside words of $e_\ell$,
\item generator replacement, such as $e_1 \to e_1'$,
\end{itemize}
which corresponds precisely to what the superscript ${\rm R}$ means in \cite{Alday:2025bjp}, and in \sec{sec:intro}.
As discussed in \app{app:Equivalence}, the ring ${\cal R}^\vee$ (which is obtained by performing the two last replacements in the elements of the original ring~${\cal R}$) should not be viewed as a separate algebraic structure --- but rather a nontrivial reorganization within the same ring ${\cal R}$.
For instance, $e_1'$ is not a separate entity but may be expanded as $e_1' = e_1 + \text{higher-order terms}$ via the original generators.

With this notion of duality in mind, let us consider the following integrand
\begin{equation}
{\cal I}^\vee(\bar{z};e'_\ell)\,\psi(\bar{z};e'_\ell) .
\label{eq:dual_system_I}
\end{equation}
For four-point open-string building blocks in AdS, we have $e'_\ell = \{e_0, e_1'\}$ and
\begin{equation}
{\cal I}^\vee(\bar{z};e_0,e_1') = \bar{z}^s(1-\bar{z})^t\,L^{\rm R}(e_0,e_1';\bar{z}) ,
\label{eq:AdS_open_building_block_dual}
\end{equation}
where the superscript in $L^{\rm R}(e_0,e_1';\bar{z})$ stands for word reversal in the MPL generating function~\eqref{eq:GenFunctionL}.
The new generator $e_1'$ is tied to $e_0$ and $e_1$ via a nontrivial relation~\eqref{eq:GeneratorEquation} \cite{Brown:2004ugm}.
This dual MPL generating function satisfies
\begin{equation}
dL^{\rm R}(e_0,e_1';\bar{z}) = L^{\rm R}(e_0,e_1';\bar{z}) \bigg[\frac{e_0}{\bar{z}}+\frac{e_1'}{\bar{z}-1}\bigg] d\bar{z} .
\end{equation}
Combining this with the similar differential equation satisfied by the conjugated Koba-Nielsen factor, we obtain the dual twist $\tau^\vee(\bar{z};e'_\ell)$:
\begin{equation}
d{\cal I}^\vee(\bar{z};e'_\ell) = {\cal I}^\vee(\bar{z};e'_\ell)\bigg[\frac{s+e_0}{\bar{z}}+\frac{t+ e_1'}{\bar{z}-1}\bigg]d\bar{z} =: {\cal I}^\vee(\bar{z};e'_\ell)\,\tau^\vee(\bar{z};e'_\ell) .
\label{eq:DLSystem}
\end{equation}
In other words, the dual twist is $\tau^\vee:=({\cal I}^\vee)^{-1}\!\times d{\cal I}^\vee$, which is the opposite order to our previous convention~\eqref{eq:NCdiff}.
This is consistent with the reverse ordering of the dual multivalued function and the dual form in \eqn{eq:dual_system_I} as compared to \eqn{eq:Int_WI}.
In order to define the dual connection, we can again consider a single patch with a particular branch of~${\cal I}^\vee$ and act with the usual exterior derivative on the product antiholomorphic form:
$d({\cal I}^\vee\psi)={\cal I}^\vee\big[d\psi + ({\cal I}^\vee)^{-1} d{\cal I}^\vee\wedge\psi\big]$.
In this way, we have
\begin{equation}
\nabla_{\tau^\vee} \psi(\bar{z};e'_\ell) := d\psi(\bar{z};e'_\ell) + \tau^\vee(\bar{z};e'_\ell)\wedge\psi(\bar{z};e'_\ell) .
\label{eq:TwistedDifferentialNC2}
\end{equation}
Just as above, we can check that $\nabla_{\tau^\vee}^2 =0$.

The reverse order of multiplication by the twist corresponds to the dual multivalued functions comprising a left ${\cal R}^\vee$-module.
Indeed, on chart overlaps $U_j \cap U_k$ the branches ${\cal I}^\vee_j$ and ${\cal I}^\vee_k$ satisfy the same equation $d{\cal I}^\vee = {\cal I}^\vee \tau^\vee$, the formal solution of which is $c\,{\cal I}^\vee$.
So the transition functions are left-multiplicative monodromy factors
\begin{equation}
{\cal I}^\vee_j(\bar{z};e'_\ell) =  g^\vee_{jk}(e'_\ell)\,{\cal I}^\vee_k(\bar{z};e'_\ell) , \quad \text{where} \quad g^\vee_{jk}(e'_\ell) \in {\cal R}^\vee \quad \forall z \in U_j \cap U_k ,
\label{eq:MonodromyFactorR}
\end{equation}
which satisfy the group relations $g^\vee_{ij} g^\vee_{jk} = g^\vee_{ik}$ (no summation).
Let us insist that we always pick the branches in the two local systems in accordance with each other, ${\cal L}_{\tau^\vee} = {\cal L}_\tau^\vee$, \ie
\begin{equation}
\big({\cal I}_j(z;e_\ell)\big)^\vee = {\cal I}^\vee_j(\bar{z};e'_\ell) .
\label{eq:LocalSystemsRelation}
\end{equation}

%%%%%%%%%%%%%%%%%%%%%%%%%%%%%%%%%%%%%%%%%%%%%%%%%%%
\subsection{Twisted cycles and cocycles}
\label{sec:co-cyles}
%%%%%%%%%%%%%%%%%%%%%%%%%%%%%%%%%%%%%%%%%%%%%%%%%%%

Just as in \sec{sec:deRham}, a twisted cocycle is a form $\omega(z;e_\ell)$ which is $\nabla_\tau$-closed, namely
\begin{equation}
\nabla_\tau\,\omega(z;e_\ell) = d\omega(z;e_\ell) + (-1)^p\,\omega(z;e_\ell) \wedge \tau(z;e_\ell) = 0 .
\label{eq:Tcocy}
\end{equation}
For a form $\omega(z;e_\ell)$ that is a solution of the above, the left multiplication $c\,\omega$ by any constant $c(e_\ell)\in{\cal R}$ naturally gives another solution, whereas the right multiplication $\omega\,c$ does not.
Therefore, we see that twisted cocycles are naturally endowed with a left-module structure.
Similarly, the dual twisted cocycles are defined as $\nabla_{\tau^\vee}$-closed forms, \ie
\begin{equation}
\nabla_{\tau^\vee}\psi(\bar{z};e'_\ell) = d\psi(\bar{z};e'_\ell) +  \tau^\vee(\bar{z};e'_\ell) \wedge \psi(\bar{z};e'_\ell) = 0 .
\label{eq:TcocyR}
\end{equation}
The situation is now opposite: if $\psi(\bar{z};e'_\ell)$ is a solution to the above, then $\psi(\bar{z};e'_\ell)\,c(e'_\ell)$ is a solution as well.
Consequently, dual cocycles have a right-module structure.
To summarize, the pattern is:
\begin{itemize}
\item twisted cocycles with respect to $\nabla_\tau$ form a left ${\cal R}$-module of differential forms satisfying $\nabla_\tau\omega(z;e_\ell)=0$,
\item twisted cocycles with respect to $\nabla_{\tau^\vee}$ form a right ${\cal R}^\vee$-module differential forms satisfying $\nabla_{\tau^\vee}\psi(\bar{z};e'_\ell)=0$.
\end{itemize}

Since both twisted connections satisfy $\nabla_\tau^2 = \nabla^2_{\tau^\vee} =0$, any $p$-form that satisfies \eqn{eq:Tcocy} or \eqref{eq:TcocyR} is still a solution if it is shifted by an exact $p$-form: $\omega \to \omega + \nabla_\tau \eta$ or $\psi \to \psi + \nabla_{\tau^\vee} \eta$, respectively.
This naturally defines the twisted cohomology classes $[\omega] \in H^p({\cal M},{\cal R},\nabla_\tau)$ and $[\psi] \in H^p({\cal M},{\cal R},\nabla_{\tau^\vee})$, where the cohomology groups are as usual
\begin{equation}
H^p({\cal M},{\cal R},\nabla_\tau) := \frac{{\rm Ker}\,\nabla_\tau}{{\rm Im}\,\nabla_\tau}\,\bigg|_{\Omega^{p,0}({\cal M},{\cal R})} , \qquad
H^p({\cal M},{\cal R},\nabla_{\tau^\vee}) := \frac{{\rm Ker}\,\,\nabla_{\tau^\vee}}{{\rm Im}\,\nabla_{\tau^\vee}}\,\bigg|_{\Omega^{0,p}({\cal M},{\cal R})} .
\label{eq:CohomologyGroups}
\end{equation}
Importantly, these cohomologies constitute left and right ${\cal R}$-modules, respectively.

On the homology side, we need to specify the notion of the twisted boundary operator~\eqref{eq:TwistedBoundarySimplex} in the noncommutative setting.
Following \sec{sec:Twisted}, we first define twisted chains as the usual topological $p$-dimensional chains equipped with the local-system data, which naturally carry the branch information of the multivalued function along the chain.
For the pair of local systems under discussion and given a usual chain $C = \sum_a \sigma_a$ constructed out of simplices, the twisted chains with local ${\cal R}$-valued coefficients are
\begin{equation}
\mathscr{C} := \sum_a\,\sigma_a \otimes {\cal I}_{\sigma_a}  ~\in C_p({\cal M},{\cal R},{\cal L}_\tau) , \qquad \quad
\tilde{\mathscr{C}} := \sum_a\,\sigma \otimes{\cal I}^\vee_{\sigma_a} ~\in C_p({\cal M},{\cal R},{\cal L}_{\tau^\vee}) .
\label{eq:TpchainL}
\end{equation}
These twisted chains naturally inherit the ${\cal R}$-module structure from the local system.
Since the original local system ${\cal L}_{\tau}$ carries the right-module structure, the twisted chains~$\mathscr{C}$ also comprise a right ${\cal R}$-module,
whereas dual twisted chains $\tilde{\mathscr{C}}$ make up a left ${\cal R}^\vee$-module.\footnote{We remind the reader that, in view of the argument in \app{app:Equivalence}, we use ${\cal R}^\vee$ to refer to the same ring~${\cal R}$ but with elements expanded in terms of the dual set of generators.
}

This module structure is the only new feature,
and the twisted boundary operator is essentially the same as in \sec{sec:Twisted}.
Namely, the boundary of a $p$-dimensional chain $\mathscr{C}$ is
\begin{equation}
\partial_\tau \mathscr{C} = \partial_\tau \bigg(\sum_a\,\sigma_a \otimes {\cal I}_{\sigma_a}\!\bigg)
:= \sum_a \partial \sigma_a \otimes {\cal I}_{\partial \sigma_a}
 = \sum_a \sum_{j=0}^p (-1)^j \hat{\sigma}_{a,\hat{j}} \otimes\,{\cal I}_{\hat{\sigma}_{a,\hat{j}}} ,
\label{eq:TwistedBoundary}
\end{equation}
where $\partial$ is the usual boundary operator, $\sigma_a$ stands for one of its constituent $p$-simplex, and $\hat{\sigma}_{a,\hat{j}}$ represents the $(p{-}1)$-simplex on its boundary obtained by deletion of the $j$th vertex.
A more explicit rendition of this formula is given in \eqn{eq:TwistedBoundarySimplex}.
Replacing ${\cal I}$ with ${\cal I}^\vee$ above computes the twisted boundary of the dual chains $\tilde{\mathscr{C}}$. For later clarity, we denote the dual boundary operator as $\partial_{\tau^\vee}$.

Twisted cycles must be boundaryless in the sense of $\partial_\tau \mathscr{C} = 0$ and similarly $\partial_{\tau^\vee} \tilde{\mathscr{C}}= 0$.
To ensure this at all times, just as in the flat-space case, we usually work with topological cycles whose boundary is on the divisor ${\cal D}$ of the multivalued function ${\cal I}$.
Since the integrand vanishes on the divisor,\footnote{A subtle point in the present case is that the multivalued integrand involves MPLs, which vanish at~0 as long as the first letter is nonzero.
However, when the first letter is~0, it vanishes only after regularization.
A common scheme is the shuffle regularization, also known as the tangential base-point regularization~\cite{Goncharov:2001iea}.
}
we are guaranteed to have a twisted cycle.
Finally, $\partial^2 = 0$ implies $\partial^2_\tau=\partial^2_{\tau^\vee}=0$,
so shifting a twisted cycle by a boundary term $\mathscr{C}  \to \mathscr{C}  + \partial_\tau \mathscr{D}$ keeps its $\partial_\tau$-closedness, and similarly $\tilde{\mathscr{C}} \to \tilde{\mathscr{C}} + \partial_{\tau^\vee} \tilde{\mathscr{D}}$ does not affect $\partial_{\tau^\vee}$-closedness.
In this way, we come to the homology classes of twisted cycles for the two systems: $[\mathscr{C}] \in H_p({\cal M},{\cal R},{\cal L}_\tau)$ and  $[\tilde{\mathscr{C}}] \in H_p({\cal M},{\cal R},{\cal L}_{\tau^\vee})$, where
\begin{equation}\!\!
H_p({\cal M},{\cal R},{\cal L}_\tau) := \frac{{\rm Ker}\,\partial_\tau}{{\rm Im}\,\partial_\tau}\,\bigg|_{C_p({\cal M},{\cal R},{\cal L}_\tau)} , \qquad
H_p({\cal M},{\cal R},{\cal L}_{\tau^\vee}) := \frac{{\rm Ker}\,\partial_{\tau^\vee}}{{\rm Im}\,\partial_{\tau^\vee}}\,\bigg|_{C_p({\cal M},{\cal R},{\cal L}_{\tau^\vee})} .
\label{eq:HomologyGroups}
\end{equation}
Note that these twisted homology groups are naturally a right ${\cal R}$-module for $H_p({\cal M},{\cal R},{\cal L}_\tau)$ and a left ${\cal R}^\vee$-module for $H_p({\cal M},{\cal R},{\cal L}_{\tau^\vee})$.

Crucially, the described ${\cal R}$-module behavior of twisted cycles and cocycles is consistent from the point of view of their integral pairings being bilinear not just with respect to the field~${\cal C}$ but in fact the whole ring~${\cal R}$.
Indeed, the product $\omega\,{\cal I}$ is left-right bilinear with respect to $\omega \to c\,\omega$ and ${\cal I} \to {\cal I}\,\tilde{c}$, where $c,\tilde{c}\in{\cal R}$ are constants.
Similarly, the dual product ${\cal I}^\vee\,\psi$ is right-left bilinear with respect to $\psi \to \psi\,c$ and ${\cal I}^\vee \to \tilde{c}\,{\cal I}^\vee$.
Finally, this freedom of left- or right- multiplication by an ${\cal R}$-valued constant makes the (co)homologies~\eqref{eq:CohomologyGroups} and~\eqref{eq:HomologyGroups} infinite-dimensional just by virtue of ${\cal R}$ being so.
However, we may think of this freedom as being trivial enough for us to abuse the language and still talk of the (co)homologies as finite-dimensional (now with respect to ${\cal R}$).

%%%%%%%%%%%%%%%%%%%%%%%%%%%%%%%%%%%%%%%%%%%%%%%%%%%
\subsection{Noncommutative pairings}
\label{sec:pairingsTDR}
%%%%%%%%%%%%%%%%%%%%%%%%%%%%%%%%%%%%%%%%%%%%%%%%%%%

In this section, we describe all the pairings starting from the periods, explain their left/right module structure, and finally describe their matrix representations.
In the noncommutative context, we can restate the Stokes theorem~\eqref{eq:TwistedStokes}
\begin{equation}
\int_{C\otimes{\cal I}_C}\!\!\nabla_\tau\omega\,
 = \int_{\partial_\tau(C\otimes{\cal I}_C)}\!\omega ,
\qquad
\begin{aligned}
(C\otimes{\cal I}_C) & \in C_{p+1}({\cal M},{\cal R},{\cal L}_\tau) , \\
\omega & \in \Omega^p({\cal M},{\cal R}) .
\end{aligned}
\label{eq:TwistedStokesNC}
\end{equation}
Here the new understanding is just that the integral of a ${\cal R}$-valued form over a twisted chain must respect the chosen order of multiplication in all of the involved structures, namely
\begin{subequations}
\begin{align}
\label{eq:IntegrationNC1}
\int_{C\otimes{\cal I}_C}\!\omega & := \sum_a\!\int_{\sigma_a}\!\!\omega\,{\cal I}_{\sigma_a} , \qquad
(C\otimes{\cal I}_C) \in C_p({\cal M},{\cal R},{\cal L}_\tau) ,~\quad
\omega \in \Omega^{p,0}({\cal M},{\cal R}) , \\
\label{eq:IntegrationNC2}
\int_{C\otimes{\cal I}_C^\vee}\!\psi & := \sum_a\!\int_{\sigma_a}\!\!{\cal I}^\vee_{\sigma_a} \psi ,\,\qquad
(C\otimes{\cal I}_C^\vee) \in C_p({\cal M},{\cal R},{\cal L}_{\tau^\vee}\!) , \quad
\omega \in \Omega^{0,p}({\cal M},{\cal R}) .
\end{align} \label{eq:IntegrationNC}%
\end{subequations}
In particular, the Stokes theorem~\eqref{eq:TwistedStokesNC} used for integrals of the type~\eqref{eq:IntegrationNC2} involves $\nabla_{\tau^\vee}$, as opposed to $\nabla_\tau$, the twist part of which acts on forms via different ${\cal R}$-multiplication, \cf \eqns{eq:TwistedDifferentialNC1}{eq:TwistedDifferentialNC2}.
In other respects, the theorem is the same, and the proof is identical to \eqn{eq:TwistedStokesProof} term by term in the ${\cal R}$-generator expansion.

So far, we have been considering $p$-dimensional chains and forms for generic $p<m$.
From now onward we will, however, focus on top (anti)holomorphic forms, \ie $m$-forms, for twisted cocycles and $m$-dimensional twisted cycles.

%%%%%%%%%%%%%%%%%%%%%%%%%%%%%%%%%%%%%%%%%%%%%%%%%%%
\subsubsection{Period pairing and its dual}
%%%%%%%%%%%%%%%%%%%%%%%%%%%%%%%%%%%%%%%%%%%%%%%%%%%

Recall that a period is a pairing between a twisted cohomology class and a twisted homology class.
In our notation, $\omega$ is a representative of a class $[\omega]$ of $\nabla_\tau$-closed ${\cal R}$-valued top holomorphic forms in the twisted cohomology $H^m({\cal M},{\cal R},\nabla_\tau)$, which has the left-module structure, and we use $\psi$ for representatives of classes in the right-module cohomology $H^m({\cal M},{\cal R},\nabla_{\tau^\vee})$.
Moreover, let $[\mathscr{C}]$ be a class of twisted $m$-cycles with ${\cal R}$-valued coefficients in the right-module homology $H_m({\cal M},{\cal R},{\cal L}_\tau)$.
The holomorphic period pairing is then
\begin{equation}
\text{period}: H^m({\cal M},{\cal R},\nabla_\tau) \times  H_m({\cal M},{\cal R},{\cal L}_\tau) \to {\cal R} :~~
\braket{[\omega],[\mathscr{C}]}_\tau := \int_{C\otimes{\cal I}_C}\!\omega
 = \sum_a\!\int_{\sigma_a}\!\!\omega\,{\cal I}_{\sigma_a} ,
\label{eq:PeriodNC}
\end{equation}
which we explicitly order in a different way from \eqn{eq:Period}, so as to mimic the noncommutative multiplication structure of the integral~\eqref{eq:IntegrationNC1}.
The invariance of this pairing is proven via \eqn{eq:PDconsistencyTwisted} but relying on the noncommutative Stokes theorem~\eqref{eq:TwistedStokesNC}.

Similarly, we define the dual (antiholomorphic) period pairing by swapping: integrate a twisted cocycle representative~$\psi$ of a class $[\psi]$ in the right-module cohomology $H^m({\cal M},{\cal R},\nabla_{\tau^\vee}\!)$ over a twisted cycle representative~$\mathscr{C}$ of a class $[\mathscr{C}]$ in the left-module homology $H_m({\cal M},{\cal R},{\cal L}_{\tau^\vee}\!)$ of the dual system.
This means
\begin{equation}\!
\substack{\text{\normalsize dual}\\\text{\normalsize period}}: H_m({\cal M},{\cal R},{\cal L}_{\tau^\vee}\!) \times H^m({\cal M},{\cal R},\nabla_{\tau^\vee}\!) \to {\cal R} :~~
\braket{[\mathscr{C}],[\psi]}_{\tau^\vee} := \int_{C\otimes{\cal I}_C^\vee}\!\psi = \sum_a\!\int_{\sigma_a}\!\!{\cal I}^\vee_{\sigma_a} \psi ,
\label{eq:PeriodDualNC}
\end{equation}
which now even looks exactly like its commutative version~\eqref{eq:PeriodDual}.
Again, since the invariance with respect to shifts by exact (co)cycles follows immediately from Stokes' theorem~\eqref{eq:TwistedStokesNC}, we will henceforth permit ourselves to drop the square brackets in the pairings.

All four (co)homologies are finite-dimensional (with respect to ${\cal R}$).
So once we pick bases~$\{\mathscr{E}^j=E^j\otimes{\cal I}_{E^j}\}$ and $\{\tilde{\mathscr{E}}_j=\tilde{E}_j\otimes{\cal I}_{\tilde{E}_j}^\vee\}$ for twisted cycles, as well as $\{\varepsilon_j\}$ and $\{\tilde\varepsilon^j\}$ for holomorphic and antiholomorphic twisted cocycles, respectively, we may encode the period pairings into two ${\cal R}$-valued period matrices:
\begin{subequations}
\begin{align}
\label{eq:PeriodMatrixNC}
P_k{}^j & := \braket{\varepsilon_k,\mathscr{E}^j}_\tau
 = \int_{E^j}\!\varepsilon_k\,{\cal I}_{E^j}
\qquad\:\!\Rightarrow~\:\qquad
\braket{\omega,\mathscr{C}}_\tau = [\omega]^k  P_k{}^j  [\mathscr{C}]_j , \\
\label{eq:PeriodMatrixDualNC}
\widetilde{P}_j{}^k & := \braket{\tilde{\mathscr{E}}_j,\tilde\varepsilon^k}_{\tau^\vee}
 = \int_{\tilde{E}_j}\!\!{\cal I}_{\tilde{E}_j}^\vee\,\tilde\varepsilon^k
\qquad \Rightarrow \qquad
\braket{\tilde{\mathscr{C}},\psi}_{\tau^\vee} = [\tilde{\mathscr{C}}]^j \widetilde{P}_j{}^k [\psi]_k .
\end{align} \label{eq:PeriodMatrices}%
\end{subequations}
Importantly, the expansion coefficients here are ${\cal R}$-valued constants, so the order of multiplication is fixed precisely as shown above.
Note that the above order is entirely consistent with the index ordering conventions introduced in \sec{sec:Dualities}.
In particular, the (co)cycle basis transformations, which are now ${\cal R}$-valued, are still described by \eqn{eq:PeriodMatrixTransform}.

%%%%%%%%%%%%%%%%%%%%%%%%%%%%%%%%%%%%%%%%%%%%%%%%%%%
\subsubsection{Poincar\'e duality}
\label{sec:PoincareDuality}
%%%%%%%%%%%%%%%%%%%%%%%%%%%%%%%%%%%%%%%%%%%%%%%%%%%

On orientiable manifolds, we may generalize \eqn{eq:CohomologyPairing} to a noncommutative cocycle pairing:
$H^m({\cal M},{\cal R},\nabla_\tau) \times H^m({\cal M},{\cal R},\nabla_{\tau^\vee}\!) \to {\cal R}$, defined by
\begin{equation}
\bbraket{[\omega],[\psi]}_{\cal M} := \int_{\cal M}\!\omega(z;e_\ell)\,{\cal I}(z;e_\ell)\,\wedge\,{\cal I}^\vee(\bar{z};e'_\ell)\,\psi(\bar{z};e'_\ell) ,
\label{eq:CocyclePairing}
\end{equation}
where we assume that the product ${\cal I}(z;e_\ell)\,{\cal I}^\vee(\bar{z};e'_\ell)$ is single-valued, so the branch may be unspecified.
At four points, this indeed follows from Brown's~\cite{Brown:2004ugm} single-valued map~\eqref{eq:SingleValuedMap}.
The invariance of the pairing~\eqref{eq:CocyclePairing} with respect to shifts by twisted-exact forms can be proven by a straightforward adjustment of \eqn{eq:CohomologyPairingProof} to the noncommutative setting.

\begin{subequations}
Now we again have the situation, where three nondegenerate bilinear pairings provide the basis for Poincar\'e duality.
Namely, we have four modules, each linear with respect to left or right multiplication by ${\cal R}$-valued numbers, such that
$H^m({\cal M},{\cal R},\nabla_{\tau^\vee}\!)$ is dual to both 
$H_m({\cal M},{\cal R},{\cal L}_{\tau^\vee})$ and $H^m({\cal M},{\cal R},\nabla_\tau)$ at the same time, meaning the latter left ${\cal R}$-modules are isomorphic to each other:
\begin{equation}
H_m({\cal M},{\cal R},{\cal L}_{\tau^\vee}\!) \leftrightarrow H^m({\cal M},{\cal R},\!\nabla_\tau)\!:\!\quad
\braket{\tilde{\mathscr{C}}(\omega),\psi}_{\tau^\vee} = \bbraket{\omega(\tilde{\mathscr{C}}),\psi}_{\cal M}\!\quad
\forall [\psi]\!\in\!H^m({\cal M},{\cal R},\!\nabla_{\tau^\vee}\!) .
\label{eq:PoincareDualityNC1}
\end{equation}
The same goes for the right ${\cal R}$-modules $H_m({\cal M},{\cal R},{\cal L}_\tau)$ and $H^m({\cal M},{\cal R},\nabla_{\tau^\vee}\!)$, which are isomorphic because of their simultaneous duality to $H^m({\cal M},{\cal R},\nabla_\tau)$:
\begin{equation}
H_m({\cal M},{\cal R},{\cal L}_\tau) \leftrightarrow H^m({\cal M},{\cal R},\nabla_{\tau^\vee}\!)\!: \quad
\braket{\omega,\mathscr{C}(\psi)}_\tau = \bbraket{\omega,\psi(\mathscr{C})}_{\cal M} \quad
\forall [\omega]\!\in\!H^m({\cal M},{\cal R},\nabla_\tau) .
\label{eq:PoincareDualityNC2}
\end{equation} \label{eq:PoincareDualityNC}%
\end{subequations}

Let us try to illustrate these claims with a heuristic implementation.
For a given representative of $[C\otimes{\cal I}_C] \in H_m({\cal M},{\cal R},{\cal L}_\tau)$, we propose the following singular $m$-form as its Poincar\'e dual:
\begin{equation}
{\rm PD}[C\otimes{\cal I}_C](\bar{z};e'_\ell)
 = \big({\cal I}_C^\vee(\bar{z};e'_\ell)\big)^{-1} \tilde{\delta}_C(\bar{z})\,d\bar{z}_1\wedge d\bar{z}_2 \wedge\cdots\wedge d\bar{z}_m ,
\label{eq:PoincareDualNC2}
\end{equation}
where $\tilde{\delta}_C$ is a distribution similar to the Dirac delta function.
Namely, it is defined by
\begin{equation}
\int_{\cal M}\!\omega(z) \wedge \tilde{\delta}_C(\bar{z})\,d\bar{z}_1 \wedge d\bar{z}_2 \wedge\cdots\wedge d\bar{z}_m
 = \int_C\!\omega(z) \qquad
\forall \omega \in \Omega^{m,0}({\cal M}) .
\label{eq:DeltaAntihol}
\end{equation}
Here we make an admittedly bold assumption that such a function exists and may be viewed as antiholomorphic, in which case the whole expression~\eqref{eq:PoincareDualNC2} is indeed a top antiholomorphic form and thus vanishes identically when acted upon with~$\nabla_{\tau^\vee}$, \ie it is a formal (dual) twisted cocycle.
It is easy to see that this construction obeys the definition~\eqref{eq:PoincareDualityNC2}:
\begin{align} &
\bbraket{\omega,{\rm PD}[\mathscr{C}]}_{\cal M}
 =\!\int_{\cal M}\!\omega(z;e_\ell)\,{\cal I}(z;e_\ell)\,\wedge\,{\cal I}^\vee(\bar{z};e'_\ell)\,{\rm PD}[C\otimes{\cal I}_C](\bar{z};e'_\ell) \\ &\!
 =\!\int_{\cal M}\!\!\!\omega\,{\cal I}\,{\cal I}^\vee ({\cal I}^\vee_C)^{-1} \wedge \tilde{\delta}_C d\bar{z}_1 \wedge\cdots\wedge d\bar{z}_m
 =\!\int_{\cal M}\!\!\!\omega\,{\cal I}_C \wedge \tilde{\delta}_C d\bar{z}_1 \wedge\cdots\wedge d\bar{z}_m
 =\!\int_C\!\!\omega\,{\cal I}_C
 = \braket{\omega,\mathscr{C}}_\tau , \nn
\end{align}
where we have used that the single-valued product ${\cal I}\,{\cal I}^\vee\!={\cal I}_C {\cal I}^\vee_C$ may always be specialized to a specific branch thanks to our assumption~\eqref{eq:LocalSystemsRelation}.

Similarly, let us introduce the conjugated Poincar\'e-dual implementation:
\begin{equation}
{\rm PD}[C\otimes{\cal I}_C^\vee](z;e_\ell)
 = \big({\cal I}_C(z;e_\ell)\big)^{-1} \delta_C(z)\,dz_1\wedge dz_2 \wedge\cdots\wedge dz_m ,
\label{eq:PoincareDualNC1}
\end{equation}
where now we assume a holomorphic realization of the delta function obeying
\begin{equation}
\int_{\cal M}\!\delta_C(z)\,dz_1 \wedge dz_2 \wedge\cdots\wedge dz_m \wedge \psi(\bar{z})
 = \int_C\!\psi(\bar{z}) \qquad
\forall \psi \in \Omega^{0,m}({\cal M}) .
\label{eq:DeltaHol}
\end{equation}
Then we have the duality~\eqref{eq:PoincareDualityNC1}, as intended,
\begin{align} &
\bbraket{{\rm PD}[\tilde{\mathscr{C}}],\psi}_{\cal M}
 =\!\int_{\cal M}\!\!{\rm PD}[C\otimes{\cal I}_C^\vee](z;e_\ell)\,{\cal I}(z;e_\ell)\,\wedge\,{\cal I}^\vee(\bar{z};e'_\ell)\,\psi(\bar{z};e'_\ell) \\ &\!
 =\!\int_{\cal M}\!\!\!{\cal I}_C^{-1} \delta_C dz_1 \wedge\cdots\wedge dz_m \wedge {\cal I}\,{\cal I}^\vee \psi
 =\!\int_{\cal M}\!\!\!\delta_C dz_1 \wedge\cdots\wedge dz_m \wedge {\cal I}_C^\vee \psi
 =\!\int_C\!\!{\cal I}_C^\vee \psi
 = \braket{\tilde{\mathscr{C}}(\omega),\psi}_{\tau^\vee} . \nn
\end{align}

%%%%%%%%%%%%%%%%%%%%%%%%%%%%%%%%%%%%%%%%%%%%%%%%%%%
\subsubsection{Intersection pairing}
\label{sec:IntersectionPairing}
%%%%%%%%%%%%%%%%%%%%%%%%%%%%%%%%%%%%%%%%%%%%%%%%%%%

Let us now use the above Poincar\`e duals to introduce the remaining twisted-cycle pairing:
\begin{equation}
H_m({\cal M},{\cal R},{\cal L}_{\tau^\vee}\!) \times H_m({\cal M},{\cal R},{\cal L}_\tau) \to {\cal R}: \quad
\braket{[\tilde{\mathscr{C}}], [\mathscr{C}]}_{\cal M}
 := \bbraket{{\rm PD}[\tilde{\mathscr{C}}],{\rm PD}[\mathscr{C}]}_{\cal M} .
\end{equation}
We then compute
\begin{align}
\braket{[\tilde{C}\otimes{\cal I}_{\tilde C}^\vee], [C\otimes{\cal I}_C]}_{\cal M} &
 = \int_{\cal M}\!\!{\rm PD}[\tilde{C}\otimes{\cal I}_{\tilde C}^\vee](z;e_\ell)\,{\cal I}(z;e_\ell)\,\wedge\,{\cal I}^\vee(\bar{z};e'_\ell)\,{\rm PD}[C\otimes{\cal I}_C](\bar{z};e'_\ell) \nn \\ &
 = \int_{\cal M}\!\!{\cal I}_{\tilde C}^{-1} \delta_{\tilde C} dz_1 \wedge\cdots\wedge dz_m\,{\cal I} \wedge {\cal I}^\vee ({\cal I}^\vee_C)^{-1}\!\wedge \tilde{\delta}_C d\bar{z}_1 \wedge\cdots\wedge d\bar{z}_m \nn \\ &
 = \int_{\cal M}\!\!\!\delta_{\tilde C} dz_1\!\wedge\!\cdots\!\wedge\!dz_m\!\wedge {\cal I}_{\tilde C}^{-1} {\cal I} {\cal I}^\vee ({\cal I} {\cal I}^\vee)^{-1} {\cal I} {\cal I}^\vee ({\cal I}^\vee_C)^{-1} \tilde{\delta}_C d\bar{z}_1\!\wedge\!\cdots\!\wedge\!d\bar{z}_m \nn \\ &
 = \int_{\cal M}\!\!\!\delta_{\tilde C} dz_1 \wedge\cdots\wedge dz_m\,{\cal I}^\vee_{\tilde C} ({\cal I} {\cal I}^\vee)^{-1} {\cal I}_C \tilde{\delta}_C d\bar{z}_1 \wedge\cdots\wedge d\bar{z}_m \nn \\ &
 =\!\sum_{z \in \tilde{C} \cap C}\!\!\pm\,{\cal I}^{\vee}_{\tilde C}\,({\cal I}\,{\cal I}^\vee)^{-1}\,{\cal I}_C ,
\label{eq:IntersectionNumberNC}
\end{align}
where in the last step we used the defining properties of the delta functions~\eqref{eq:DeltaAntihol} and~\eqref{eq:DeltaHol}.
They localized the volume integral on their support --- the intersection points between the two closed contours~$\tilde{C}$ and $C$.
\Eqn{eq:IntersectionNumberNC} is clearly a noncommutative generalization of the intersection-pairing formula~\eqref{eq:IntersectionNumber}.
Again, we have intersection points that are weighted by the orientation-dependent signs times the monodromy factors of the multivalued function depending on the branches chosen by the two contours.
An important point to note is that the intersection pairing requires that at least one of the twisted cycles must be compact~\cite{yoshida1997hypergeometric,Aomoto:2011ggg}.

%%%%%%%%%%%%%%%%%%%%%%%%%%%%%%%%%%%%%%%%%%%%%%%%%%%
\subsection{Noncommutative twisted period relations}
\label{sec:INumber}
%%%%%%%%%%%%%%%%%%%%%%%%%%%%%%%%%%%%%%%%%%%%%%%%%%%

In this section, we derive the twisted period relations for our ${\cal R}$-module (co)homologies.
As shown in \fig{fig:Dualities}, all the bilinear pairings we encountered combine an element of a left ${\cal R}$-module with an element of a right ${\cal R}$-module.
Poincar\'e duality, on the other hand, relates two left modules in \eqn{eq:PoincareDualityNC1}, as well two right modules in \eqn{eq:PoincareDualityNC2}.

We have already introduced the ${\cal R}$-valued period matrices~\eqref{eq:PeriodMatrices} obtained by expanding the four (co)homologies in terms of bases~$\{\mathscr{E}^j\}$ and $\{\tilde{\mathscr{E}}_j\}$ for twisted cycles, as well as $\{\varepsilon_j\}$ and $\{\tilde\varepsilon^j\}$ for holomorphic and antiholomorphic twisted cocycles.
Similarly, we can express the twisted-cocycle pairing~\eqref{eq:CocyclePairing} via another ${\cal R}$-valued matrix:
\begin{equation}
F_j{}^k := \bbraket{\varepsilon_j,\tilde\varepsilon^k}_{\cal M}
 = \int_{\cal M}\!\varepsilon_j\,{\cal I}\wedge{\cal I}^\vee \tilde\varepsilon^k \qquad \Rightarrow \qquad
\bbraket{\omega,\psi}_{\cal M} := [\omega]^j F_j{}^k [\psi]_k .
\label{eq:CocycleMatrixNC}
\end{equation}
Poincar\'e duality may then be expressed by choosing the basis cocycles in \eqn{eq:PoincareDualityNC}
\begin{subequations}
\begin{align}
\label{eq:TowardsPoincareDualityMatrixNC1}
\big[{\rm PD}[\varepsilon_j]\big]^k \widetilde{P}_k{}^l [\psi]_l = \braket{{\rm PD}[\varepsilon_j],\psi}_{\tau^\vee} &
 = \bbraket{\varepsilon_j,\psi}_{\cal M} = F_j{}^l [\psi]_l , \\
\label{eq:TowardsPoincareDualityMatrixNC2}
[\omega]^l P_l{}^k \big[{\rm PD}[\tilde\varepsilon^j]\big]_k = \braket{\omega,{\rm PD}[\tilde\varepsilon^j]}_\tau &
 = \bbraket{\omega,\tilde\varepsilon^j}_{\cal M} = [\omega]^l F_l{}^j .
\end{align} \label{eq:TowardsPoincareDualityMatricesNC}%
\end{subequations}
From this, we immediately obtain
\begin{subequations}
\begin{equation}
\big[{\rm PD}[\varepsilon_j]\big]^k = F_j{}^l \widetilde{P}^{-1}{}_l{}^k ,
\qquad \quad
\big[{\rm PD}[\tilde\varepsilon^j]\big]_k = P^{-1}{}_k{}^l F_l{}^j .
\label{eq:PoincareDualityMatricesNC1}
\end{equation}
This is just a minor rewriting of \eqn{eq:PoincareDualityMatrices}, which again justifies our choice of conventions back in the commutative exposition of \sec{sec:Dualities}.
The inverse relations are
\begin{equation}
\big[{\rm PD}[\tilde{\mathscr{E}}_j]\big]^k = \widetilde{P}_j{}^l F^{-1}{}_l{}^k ,
\qquad \quad
\big[{\rm PD}[\mathscr{E}^j]\big]_k = F^{-1}{}_k{}^l P_l{}^j  .
\label{eq:PoincareDualityMatricesNC2}
\end{equation} \label{eq:PoincareDualityMatricesNC}%
\end{subequations}

The ${\cal R}$-valued intersection-number matrix may also be introduced in complete analogy to \eqn{eq:IntersectionNumberMatrix}.
The full computation is
\begin{equation}
\begin{aligned}
I_j{}^k := &\,\braket{\tilde{\mathscr{E}}_j,\mathscr{E}^k}_{\cal M}
 = \bbraket{{\rm PD}[\tilde{\mathscr{E}}_j],{\rm PD}[\mathscr{E}^k]}_{\cal M}
 = \big[{\rm PD}[\tilde{\mathscr{E}}_j]\big]^i\,F_i{}^l\,\big[{\rm PD}[\mathscr{E}^k]\big]_l \\
 = &\,[\tilde{\mathscr{E}}_j]^a \widetilde{P}_a{}^b F^{-1}{}_b{}^i F_i{}^l F^{-1}{}_l{}^r P_r{}^s [\mathscr{E}^k]_s
 = \widetilde{P}_j{}^i F^{-1}{}_i{}^l P_l{}^k .
\label{eq:IntersectionNumberMatrixNC}
\end{aligned}
\end{equation}
Therefore, the twisted period relations have the same form as in \eqn{eq:PeriodRelations}, namely
\begin{equation}
I = \widetilde{P} F^{-1} P \qquad \Leftrightarrow \qquad F = P I^{-1} \widetilde{P} .
\label{eq:PeriodRelationsNC}
\end{equation}
The explicit formula for the ${\cal R}$-valued intersection number of two basis twisted cycles $\tilde{\mathscr{E}}_j=\tilde{E}_j\otimes{\cal I}_{\tilde{E}_j}^\vee$ and $\mathscr{E}^k=E^k\otimes{\cal I}_{E^k}$ is
\begin{equation}
I_j{}^k =\!\sum_{z \in \tilde{E}_j \cap E^k}\!\!\pm\,{\cal I}_{\tilde{E}_j}^\vee\,({\cal I}\,{\cal I}^\vee)^{-1}\,{\cal I}_{E^k} .
\label{eq:Intersections}
\end{equation}
Note that, since this pairing is actually between two equivalence classes, we are allowed to deform the contours so that the deformation is a boundary of an enclosed region, because this does not change the equivalence class and thus not the intersection number either.

%%%%%%%%%%%%%%%%%%%%%%%%%%%%%%%%%%%%%%%%%%%%%%%%%%%
\subsection{AdS double copy at 4 points}
\label{sec:DC4pt}
%%%%%%%%%%%%%%%%%%%%%%%%%%%%%%%%%%%%%%%%%%%%%%%%%%%

In this section, we finally provide the proof of the AdS building-block double-copy relation~\eqref{eq:AdS_double_copy}.
For the four-point AdS open-string integral, the homology group has only one basis element, as already mentioned in \eqn{eq:FlatSpaceHomologyTwisted}.
For the corresponding twisted cohomology, we use the following ``Parke-Taylor'' factor as the basis,
\begin{equation}
\omega(z;e_\ell)={\rm PT}(z)= \frac{dz}{z(1-z)} = d\log\frac{z}{1-z} ,
\label{eq:PTcocycle}
\end{equation}
for which the flat-space open-string integral is Veneziano's Euler beta function~\eqref{eq:VenezianoAmplitudePeriod}.
This rational function is a particular case of the Parke-Taylor-like cyclic denominator $1/\prod_j (z_j{-}z_{j+1})$, which is ubiquitous in massless-particle scattering~\cite{Cachazo:2013hca,Cachazo:2013iea}, where $z_j$ stand for punctures of the two-dimensional worldsheet.
We have already specialized to the single color ordering of these punctures: $z_1<z_2<z_3<z_4$.
Moreover, we have gauge-fixed the locations of three of the punctures to $z_1=0$; $z_3=1$; $z_4=\infty$ and renamed $z_2=:z$.
Therefore, the form of the twisted cocycle~\eqref{eq:PTcocycle} is a special case in our construction of the twisted de Rham theory for noncommutative case.
In fact, since it is independent of~$e_\ell$, it commutes with a generic ${\cal R}$-valued integrand.
Consequently, one may wonder if it was even necessary to start with a generic ${\cal R}$-valued form $\omega(z;e_\ell)$.
However, despite using a special $\mathbb{C}$-valued forms in this section, we insist that more general twisted cocycles are solutions of $\nabla_\tau \omega(z;e_\ell)$ and hence by construction have the freedom of left multiplication by ring-valued constants.
As soon as one does that, the twisted cocycles stop commuting with other objects and, as demonstrated in previous sections, one has to respect the prescribed ordering of the period pairing.

For the dual period, we simply choose
\begin{equation}
\psi(\bar{z};e'_\ell) = {\rm PT}^\vee(\bar{z})= \frac{d\bar{z}}{\bar{z}(1-\bar{z})} = d\log\frac{\bar{z}}{1-\bar{z}} ,
\label{eq:PTcocycleDual}
\end{equation}
where the $\vee$ operation, which we also call reversal, reduces to complex conjugation, because it has no explicit dependence on the generators $e'_\ell$.
Given that the dimension of the twisted homology group is same as the twisted cohomology group, we use
\begin{equation}
\mathscr{C} = (0,1) \otimes {\cal I}(z;e_\ell) ,
\qquad \text{where} \qquad
{\cal I}(z;e_\ell) := z^s (1-z)^t L(e_\ell;z) ,
\label{eq:TwistedCycle1}
\end{equation}
as our single basis element for twisted cycles with ${\cal R}$-valued coefficients.
Similarly, for the dual cycle we have
\begin{equation}
\tilde{\mathscr{C}} = (0,1) \otimes {\cal I}^\vee(\bar{z};e'_\ell) ,
\qquad \text{where} \qquad
{\cal I}^\vee(\bar{z};e'_\ell) := \bar{z}^s (1-\bar{z})^t L^\vee(e'_\ell;\bar{z}) .
\label{eq:TwistedCycle2}
\end{equation}
Here and below, we assume that ${\cal I}$ is by default evaluated on the principal branch and suppress the branch subscripts of ${\cal I}$.
Then the period pairings defined by \eqns{eq:PeriodNC}{eq:PeriodDualNC} are
\begin{subequations}
\begin{align}
\braket{{\rm PT}(z),\mathscr{C}}_\tau & = \int_0^1\!z^{s-1}(1-z)^{t-1}L(e_\ell;z)\,dz
 = J(e_\ell;s,t) , \\
\braket{\tilde{\mathscr{C}},{\rm PT}^\vee(\bar{z})}_{\tau^\vee} & = \int_0^1\!\bar{z}^{s-1}(1-\bar{z})^{t-1}L^\vee(e'_\ell;\bar{z})\,d\bar{z}
 = J^\vee(e'_\ell;s,t) .
\end{align}
\end{subequations}
On the other hand, the cohomology pairing~\eqref{eq:CocyclePairing} between the twisted cocycle~\eqref{eq:PTcocycle} and its dual~\eqref{eq:PTcocycleDual} reads
\begin{equation}
\begin{aligned}
\bbraket{{\rm PT}(z),{\rm PT}^\vee(\bar{z})}_{\cal M} &
 = \int_{\mathbb{C}\setminus\{0,1\}}\!|z|^{2s-2} |1-z|^{2t-2}L(z;e_w)L^\vee(\bar{z};e'_\ell)\,d z \wedge d\bar{z} \\ &
 = \int_{\mathbb{C}\setminus\{0,1\}}\!|z|^{2s-2}|1-z|^{2t-2}{\cal L}(z;e_\ell)\, d z\wedge d\bar{z} \\ &
 = -2i I(e_\ell;s,t) ,
\end{aligned}
\end{equation}
where we have used the single-valued map~\eqref{eq:SingleValuedMap} from the MPL generating functions to the SVMPL generating function~\cite{Brown:2004ugm}, and $I(e_\ell;s,t)$ is the AdS-curvature-corrected four-point closed-string integral~\eqref{eq:closed_string_AdS} introduced in \sec{sec:intro}.
The factor of $-2i$ comes from our measure convention $dz \wedge d\bar{z} =:-2i d^2z$.

Since the (co)homologies are one-dimensional (with respect to ring ${\cal R}$), the twisted period relations~\eqref{eq:PeriodRelationsNC} are a single equation
\begin{equation}
\bbraket{{\rm PT}(z),{\rm PT}^\vee(\bar{z})}_{\cal M} = \braket{{\rm PT}(z), \mathscr{C}}_\tau \times \braket{\tilde{\mathscr{C}},\mathscr{C}}_{\cal M}^{-1} \times \braket{\tilde{\mathscr{C}},{\rm PT}^\vee(\bar{z})}_{\tau^\vee} .
\label{eq:PeriodRelationNC}
\end{equation}
The only missing ingredient in the above expression is the intersection number of the twisted cycles~\eqref{eq:TwistedCycle1} and~\eqref{eq:TwistedCycle2}.
Recall, however, that it requires at least one of the cycles to be compact~\cite{yoshida1997hypergeometric,Aomoto:2011ggg}.
This is a problem, because the open-string integrals considered above are both defined over the noncompact oriented contour $(0,1)$ due to it starting and ending on the punctures.
In order to compute their intersection number, one needs a regularization procedure, which facilitates an isomorphism between the twisted homology group that we have to a compact twisted homology group.

\begin{figure}[t]
\centering
\includegraphics[width=0.33\textwidth]{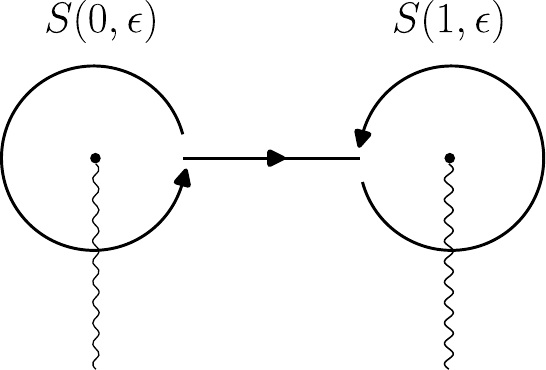}
\vspace{-5pt}
\caption{Regularized twisted cycle ${\rm reg}(0,1)$. It consists of two circles $S(j,\epsilon)$ centered at $j \in \{0,1\}$ of radius $\epsilon$ oriented anticlockwise. These are connected via a closed interval $[\epsilon, 1-\epsilon]$ along the real axis. The wiggly lines denote the branch cuts, both oriented along the negative imaginary axis.}
\label{fig:reg}
\end{figure}

%%%%%%%%%%%%%%%%%%%%%%%%%%%%%%%%%%%%%%%%%%%%%%%%%%%
\subsubsection{Regularized twisted cycle}
\label{sec:Reg01}
%%%%%%%%%%%%%%%%%%%%%%%%%%%%%%%%%%%%%%%%%%%%%%%%%%%

Recall that our twisted cycle~\eqref{eq:PTcocycleDual} has the tensor-product structure, where the local coefficient is ${\cal R}$-valued.
Moreover, since the homology group $H_1(\mathbb{C}\!\setminus\!\{0,1\},{\cal R},{\cal L}_\tau)$ is a right module, we are allowed to further multiply it by an ${\cal R}$-valued constant.
With this in mind and following \cite{mana.19941660122,Mimachi:2002gi}, we introduce the following regularization of the open-interval cycle $\mathscr{C}=(0,1)\otimes{\cal I}(z;e_\ell)$:
\begin{equation}
\text{reg}(0,1) = S(0,\epsilon) \frac{1}{e^{2\pi i s}M_0-1}\,+ [\epsilon,1-\epsilon] \otimes {\cal I}(z;e_\ell) + S(1,\epsilon) \frac{1}{1-e^{2\pi i t}M_1} .
\label{eq:REG_01}
\end{equation}
Here $[\epsilon,1-\epsilon]$ is simply the closed interval along the real axis connecting the two circles $S(0,\epsilon)$ and $S(1,\epsilon)$, see \fig{fig:reg}.
More precisely, $S(0,\epsilon)$ and $S(1,\epsilon)$ denote the twisted chains that correspond to infinitesimal circular contours $\circlearrowleft_0^\epsilon$ and $\circlearrowleft_1^\epsilon$ of radius $\epsilon$ centered at~0 and~1, respectively:
\begin{equation}
S(0,\epsilon) =\;\circlearrowleft_0^\epsilon \otimes\,{\cal I}(z;e_\ell) , \qquad \quad
S(1,\epsilon) =\;\circlearrowleft_1^\epsilon \otimes\,{\cal I}(z;e_\ell) .
\end{equation}
Finally, \eqn{eq:REG_01} has both circles multiplied on the right by explicit ${\cal R}$-valued coefficients, which are related to the following monodromy factors on the right, as per \eqn{eq:Monodromies01}:
\begin{equation}
M_0=e^{2\pi i e_0}, \qquad M_1=Z(e_1,e_0)e^{2\pi i e_1}Z(e_0,e_1) .
\end{equation}
Pictorially, one should think of the above compact cycle as follows: the closed interval as well as its intersection points with the two circles are all on the same (principal) branch of the multivalued function ${\cal I}$.
Therefore, when we move along the two circles anticlockwise, we cross the associated branch and thus pick up the monodromy factors (via right ${\cal R}$-multiplication) and end up on different sheets.
Given the multivalued function in \eqn{eq:TwistedCycle1}, there are two monodromy factors to account for: one for the Koba-Nielsen factor, see \eqn{eq:LocalSystem}, and another for the MPL generating function at~0 and~1.

An important point to note in \eqn{eq:REG_01} is that each of the two circles has an associated coefficient depending on the monodromy of ${\cal I}$ around their respective centers.
These are not arbitrary and are explicitly fixed by the requirement that $\text{reg}(0,1)$ is indeed a twisted cycle, \ie $\partial_\tau[\text{reg}(0,1)]=0$, otherwise it has nothing to do with the homology group.
To verify this claim, let us for a moment unfix the coefficients in \eqn{eq:REG_01} and assume
\begin{equation}
\text{reg}(0,1) = S(0,\epsilon)\,\alpha_1(e_\ell;s,t)
 + [\epsilon,1-\epsilon] \otimes {\cal I}(z;e_\ell)\,\alpha_2(e_\ell;s,t)
 + S(1,\epsilon)\,\alpha_3(e_\ell;s,t) .
\label{eq:REG_ref}
\end{equation}
For its twisted boundary, we derive
\begin{align}
\partial_\tau \big[\text{reg}(0,1)\big] &
 = \partial_\tau \big(\!\circlearrowleft_0^\epsilon\otimes\,{\cal I}_{\circlearrowleft_0^\epsilon}\big)\,\alpha_1
 + \big[(1-\epsilon)\otimes{\cal I}(1-\epsilon) - \epsilon\otimes{\cal I}(\epsilon)\big]\,\alpha_2
 + \partial_\tau \big(\!\circlearrowleft_1^\epsilon\otimes\,{\cal I}_{\circlearrowleft_1^\epsilon}\big)\,\alpha_3 \nn \\ &
 = \epsilon\otimes{\cal I}(\epsilon) [e^{2\pi i s}M_0\!-1]\,\alpha_1
 + \big[(1-\epsilon)\otimes{\cal I}(1-\epsilon)-\epsilon\otimes{\cal I}(\epsilon)\big]\,\alpha_2 \\ & \qquad \qquad \qquad \qquad \qquad~\:\quad
 + (1-\epsilon)\otimes{\cal I}(1-\epsilon)[e^{2\pi i t}M_1\!-1]\,\alpha_3 \nn \\ &
 = \epsilon\otimes{\cal I}(\epsilon) \big([e^{2\pi i s}M_0\!-1]\,\alpha_1 - \alpha_2 \big)
 + (1-\epsilon) \otimes {\cal I}(1-\epsilon) \big(\alpha_2 + [e^{2\pi i t}M_1\!-1]\,\alpha_3 \big) , \nn
\end{align}
where we have used the monodromy relations~\eqref{eq:TwistedNonCycles} of $z^s (1-z)^t$ enhanced by the ${\cal R}$-valued monodromies~\eqref{eq:Monodromies01} of $L(e_\ell;z)$, which necessarily multiply on the right.
This is why we chose the coefficients $\alpha_j$'s to be right multipliers, otherwise noncommutativity would create issues.
Imposing that the twisted boundary above should vanish, we obtain the following solutions for the coefficients
\begin{equation}
\alpha_1 = \frac{1}{[e^{2\pi i s}M_0\!-1]}\,\alpha_2 , \qquad \qquad
\alpha_3 =- \frac{1}{[e^{2\pi i t}M_1\!-1]}\,\alpha_2 .
\end{equation}
Picking $\alpha_2 = 1$ and substituting into \eqn{eq:REG_ref},
we recover the claimed twisted cycle~\eqref{eq:REG_01}.

%%%%%%%%%%%%%%%%%%%%%%%%%%%%%%%%%%%%%%%%%%%%%%%%%%%
\subsubsection{Intersection number}
\label{sec:IntersectionNumber}
%%%%%%%%%%%%%%%%%%%%%%%%%%%%%%%%%%%%%%%%%%%%%%%%%%%

As for the dual chain $\tilde{\mathscr{C}} = (0,1) \otimes {\cal I}^\vee(\bar{z};e'_\ell)$, we may leave it unregularized as an open interval.
We are, however, also allowed to deform it so that it is still in the same equivalence class.
Below we will consider two such deformations and demonstrate that the intersection number is indeed invariant under such a deformation. 

\begin{figure}[t]
\centering
\includegraphics[width=0.33\textwidth]{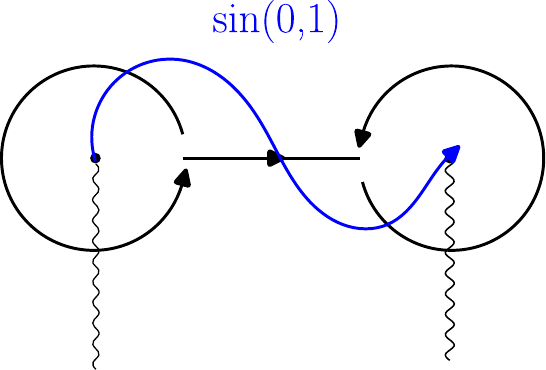}
\vspace{-5pt}
\caption{Sinusoid-shaped twisted cycle ${\rm sin}(0,1)$. The black contour is the regularized cycle ${\rm reg}(0,1)$, whereas the blue contour is the sinusoidal deformation of the real-axis interval $(0,1)$. ${\rm sin}(0,1)$ differs from $(0,1)$ by two semi-circular regions enclosed between them and is therefore in the same equivalence class.}
\label{fig:sin}
\end{figure}

The first deformation will be to a sinusoidal shape, denoted as ${\rm sin}(0,1)$, see \fig{fig:sin}.
This deformation is to avoid a continuous overlap with the closed interval $[\epsilon, 1-\epsilon]$ in the regularized cycle~\eqref{eq:REG_01}.
For this chosen deformation, the intersection number is given by
\begin{equation}
\braket{\tilde{\mathscr{C}},\mathscr{C}}_{\cal M}
 = \braket{{\rm sin}(0,1),\text{reg}(0,1)}_{\cal M}
 = \frac{(-1)}{e^{2\pi is}M_0-1} + (-1) + \frac{(+1)}{1-e^{2\pi i t}M_1} ,
\label{eq:Intersections1}
\end{equation}
where we have used the master formula~\eqref{eq:Intersections}.
Let us explain the three terms one by one, each corresponding to a geometrical intersection point of the contours.
For each crossing, the intersection number is the coefficient associated with each of the two contours at the point of intersection times an orientation factor according to
\begin{equation}
\begin{aligned}\includegraphics[width=0.12\textwidth]{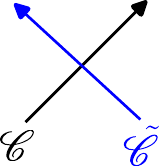}\end{aligned} ~\Rightarrow~ +1 , \qquad \qquad \quad
\begin{aligned}\includegraphics[width=0.12\textwidth]{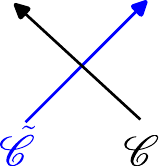}\end{aligned} ~\Rightarrow~ -1 .
\label{eq:IntersectionSigns}
\end{equation}
These signs are highlighted in \eqn{eq:Intersections1} as the numerators $\{(-1),(-1),(+1)\}$, one for each intersection point respectively.
Notice that the first and the third intersection contributions include the coefficients associated with the circles $S(0,\epsilon)$, and $S(1,\epsilon)$, respectively.

\begin{figure}[t]
\centering
\includegraphics[width=0.33\textwidth]{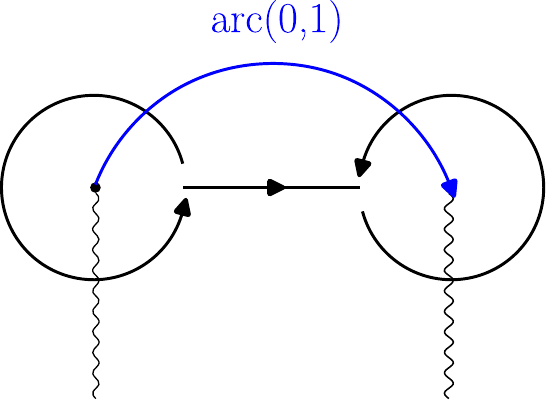}
\vspace{-5pt}
\caption{Arc-shaped twisted cycle ${\rm arc}(0,1)$. The black contour is the regularized cycle ${\rm reg}(0,1)$, whereas the blue contour is an arched deformation of the real-axis interval $(0,1)$.  ${\rm arc}(0,1)$ differs from $(0,1)$ by the big semi-circular region enclosed between them and is therefore in the same equivalence class.}
\label{fig:arc}
\end{figure}

Let us consider a second deformation shown in \fig{fig:arc}, which we denote as ${\rm arc}(0,1)$.
Now there are only two intersection points that contribute to the master formula~\eqref{eq:Intersections}, coming with signs $+1$ and $-1$.
For this deformation the intersection number is
\begin{equation}
\braket{\tilde{\mathscr{C}},\mathscr{C}}_{\cal M}
 = \braket{{\rm arc}(0,1),\text{reg}(0,1)}_{\cal M} 
 = \frac{(-1)}{e^{2\pi is}M_0-1} + \frac{(+1)}{1-e^{2\pi i t}M_1} e^{2\pi i t}M_1 .
\label{eq:Intersections2}
\end{equation}
The first intersection term is entirely analogous to that in \eqn{eq:Intersections1}.
The second intersection terms, however, is more interesting, as it comes from the arc intersecting the circle $S(1,\epsilon)=\circlearrowleft_1^\epsilon \otimes\,{\cal I}(z;e_\ell)$ after it crossed the branch cut, so $\text{reg}(0,1)$ has to pick up an additional monodromy factor $e^{2\pi i t}M_1$ on the right.
It is easy to observe that the results of \eqns{eq:Intersections1}{eq:Intersections2} coincide and are equal to
\begin{equation}
\braket{\tilde{\mathscr{C}},\mathscr{C}}_{\cal M} = 1+\frac{e^{2\pi i s}M_0}{1-e^{2\pi i s}M_0}+\frac{e^{2\pi i t}M_1}{1-e^{2\pi i t}M_1} .
\end{equation}
Putting everything back into the twisted period relation~\eqref{eq:PeriodRelationNC}, we have
\begin{equation}
\begin{aligned}
I(e_0,e_1;s,t) & = \frac{i}{2} J(e_0,e_1;s,t) \left(1+\frac{e^{2\pi i s}M_0}{1-e^{2\pi i s}M_0}+\frac{e^{2\pi i t}M_1}{1-e^{2\pi i t}M_1}\right)^{\!-1}\!J^\vee(e_0,e_1';s,t) \\ &
 = J(e_0,e_1;s,t)\,{\cal K}(e_0,e_1;s,t)\,J^\vee(e_0,e_1';s,t) ,
\end{aligned}    
\end{equation}
which completes the proof of \eqn{eq:AdS_double_copy}.

%%%%%%%%%%%%%%%%%%%%%%%%%%%%%%%%%%%%%%%%%%%%%%%%%%
\section{Conclusion and discussion}
\label{sec:outro}
%%%%%%%%%%%%%%%%%%%%%%%%%%%%%%%%%%%%%%%%%%%%%%%%%%%

In this work, we have introduced the framework of noncommutative twisted de Rham theory.
Its construction was made possible mainly by two important facts:
\begin{itemize}
\item the MPL and SVMPL generating series obey the single-valued map~\eqn{eq:SingleValuedMap}~\cite{Brown:2004ugm};
\item the monodromies of the MPL generating function are multiplicative~\cite{MINH2000217}, and its twist is single-valued.
\end{itemize}
As an application of this framework,
we have provided a derivation of the AdS double-copy relation~\eqref{eq:AdS_double_copy} between the curvature-corrected four-point Veneziano and Virasoro–Shapiro integrals~\eqref{eq:open_string_AdS} and~\eqref{eq:closed_string_AdS}.
These integrals collect the infinite tower of building blocks encoding the AdS curvature corrections both for open- and closed-string amplitudes \cite{Alday:2025bjp}.

The formulation of the noncommutative twisted de Rham theory begins with the identification of the AdS open-string generating function \eqref{eq:open_string_AdS} as a period --- a natural pairing between a twisted cycle and a twisted cocycle.
The noncommutative twist \eqref{eq:sttwist} is then defined by the multivalued function inside the twisted cycles, consisting of the Koba-Nielsen factor times the MPL generating series encoding both $\alpha'$-effects and AdS curvature corrections.
The above stated single-valued map \eqref{eq:SingleValuedMap} allows us to define the dual system, following which we have two dual periods as natural pairings between twisted cycles and cocycles.
The analogous pairings in flat space correspond to the left- and right-mover open-string amplitudes.
The closed-string generating function~\eqref{eq:closed_string_AdS} in this formalism emerges as a bilinear pairing of twisted cocycles, and the AdS double-copy kernel as the inverse of the intersection pairing of twisted cycles.
The twisted period relations~\eqref{eq:PeriodRelationsNC}, which results from Poincar\'e duality, then combine all these pairings to give the AdS double copy~\eqref{eq:AdS_double_copy}.
Therefore, this approach establishes that the AdS double-copy relations may not be merely special-function identities but instead follow from an underlying noncommutative twisted de Rham theory governing string worldsheet integrals.

Our proof places AdS double-copy relations on the same conceptual footing as the flat-space KLT relations between the open- and closed-string amplitudes.
In fact, the noncommutative curvature-deformed twisted de Rham theory constructed here reduces smoothly to the standard twisted de Rham theory, for which the twist comes solely from the Koba–Nielsen factor.
This gives hope that such an approach may provide a unified framework for understanding open– and closed-string relations across backgrounds with different curvatures.

Several open problems and directions for future research naturally emerge from this work, which will make contact with the other active research in this direction~\cite{Schlotterer:2018zce,Brown:2018omk,Brown:2019wna,Albayrak:2020fyp,Alday:2021odx,Zhou:2021gnu,Alday:2024srr,Frost:2023stm,Alday:2024xpq,Chester:2024esn,Frost:2025lre,Chen:2025cod,Baune:2025hfu}.
One immediate extension is to derive the open- and closed-string AdS double copy for four-point amplitudes,
which can be obtained from the building blocks considered here via an inner-product scheme~\cite{Alday:2025cxr}.
The authors of~\cite{Alday:2025cxr} computed the amplitudes with first-order curvature corrections, which required including all the corrections from the MPL generating series up to weight~3.
However, we believe that if an order-by-order (in curvature) double copy exists, it will be highly nontrivial simply because according to our proof the double copy for generating series was natural owing to the single-valued map~\eqref{eq:SingleValuedMap}, which only holds for the entire generating series of MPLs and SVMPLs.
Moreover, given the recurring theme of the flat-space KLT and the AdS double copy naturally emerging within twisted de Rham theory, we suspect that there should be a natural extension of our formulation that incorporates the above stated inner-product scheme and thus drives the double copy for the AdS amplitudes.

In \cite{Mizera:2016jhj}, the inverse of the flat-space KLT kernel was identified as certain $\alpha'$-corrected bi-adjoint scalar amplitudes.
In the language of twisted de Rham theory, this corresponds to the intersection number of twisted cycles.
This implies that the intersection number of twisted cycles computed in our approach could correspond to the AdS curvature-corrected building blocks for the bi-adjoint scalar amplitudes.

Another problem is to consider higher-point open- and closed-string amplitudes in AdS.
Our approach highlights the universality of twisted de Rham methods in organizing the double-copy relations among open- and closed-string amplitudes across different backgrounds and thus may suggest natural generalizations to higher points~\cite{Broedel:2013aza,Mizera:2017cqs,Mizera:2019gea,Casali:2019ihm,Britto:2021prf,Mafra:2022wml,Baune:2024uwj}, which we leave for future work.
A closely related, however farsighted goal, is the question of loop-level generalizations of the string double copy~\cite{Tourkine:2016bak,Hohenegger:2017kqy,Ochirov:2017jby,Broedel:2018izr,Casali:2020knc,Stieberger:2023nol,Bhardwaj:2023vvm,Mazloumi:2024wys,Pokraka:2025zlh}, where twisted de Rham on higher-genus moduli spaces could provide a geometric understanding of open-closed string amplitude relations beyond tree level.
Another important direction is the application of these ideas to other curved backgrounds, including de Sitter space, where the existence and form of KLT-like relations remain largely unexplored.
The hope is that there should be a far richer twisted de Rham formalism that can unify the current construction and provide new algebraic and geometric understanding to double-copy relations in general.

%%%%%%%%%%%%%%%%%%%%%%%%%%%%%%%%%%%%%%%%%%%%%%%%%%
\begin{acknowledgments}
We thank Fernando Alday, Promit Kundu, R\'emy Larue, Rodrigo Pitombo, Carlos Rodriguez and Aur\'elie Str\"omholm Sangar\'e for enlightening conversations.
HK is supported by the National Natural Science Foundation of China via NSFC grant No. 12447147.
HK and AO are supported by the Science and Technology Commission of the Shanghai
Municipality via STCSM grant No. 24ZR1450600.

\end{acknowledgments}
%%%%%%%%%%%%%%%%%%%%%%%%%%%%%%%%%%%%%%%%%%%%%%%%%%

\appendix

\section{Duality transformation as ring automorphism}
\label{app:Equivalence}

Let ${\cal R} = \widehat{\mathbb{C}\braket{e_0,e_1}}$ be the completed free associative algebra generated by the symbols $e_0$ and $e_1$.
Consider an element $Z \in \cal{R}$, which may be inverted as a formal series.
This means that the expansion of $Z$ starts with a nonzero number:
\begin{equation}
Z = c + {\cal O}(e_0, e_1) , \qquad \quad 0 \neq c \in \mathbb{C} ,
\end{equation}
where the higher-order terms contain at least one generator.
(For concreteness, $Z$ may be renormalized to have $c=1$.)
We can define a transformed generator $\tilde{e}_1$ as
\begin{equation}
\tilde{e}_1 := Z\,e_1\,Z^{-1}
 = \big(1 + {\cal O}(e_0, e_1)\big)\,e_1\,\big(1 - {\cal O}(e_0, e_1)\big)
 = e_1 + \text{higher-order terms} .
\label{eq:NewGenerator}
\end{equation}

We claim that the algebra generated by $\{e_0, \tilde{e}_1\}$ is identical to the original algebra ${\cal R}$.
In other words, the transformation~\eqref{eq:NewGenerator} is a redefinition of the generators that preserves the underlying algebraic structure.
To see this, note that the structure of \eqn{eq:NewGenerator} allows it to be inverted order by order in the length (or weight) grading of the algebra.
Consequently, one can uniquely expressed the original generator $e_1$ as a power series in $e_0$ and $\tilde{e}_1$.
Any element of ${\cal R}$ can thus also be uniquely rewritten in terms of the generators.
Formally, the map $e_0 \to e_0$ and $e_1 \to \tilde{e}_1$ extends to an algebra automorphism of ${\cal R}$. Therefore, the descriptions using the basis $\{e_0, e_1\}$ and the basis $\{e_0, \tilde{e}_1\}$ are equivalent:
\begin{equation}
{\cal R} = \widehat{\mathbb{C}\braket{e_0,e_1}}
 = \widehat{\mathbb{C}\braket{e_0, Z e_1 Z^{-1}}} .
\end{equation}

The Drinfeld associator~\eqref{eq:DrinfeldAssociator} is a specific invertible element of ${\cal R}$, which maps the two generators $e_1$ and $e_1'$ discussed in the main text to the same $\tilde{e}_1$~\cite{Brown:2004ugm}, namely
\begin{equation}
Z^\vee(e_0, e_1')\,e_1'\big(Z^\vee(e_0, e_1')\big)^{-1}
 = Z(e_1, e_0)\,e_1\,Z(e_1, e_0)^{-1} .
\label{eq:GeneratorEquation}
\end{equation}
This implies that the dual ring ${\cal R}^\vee$ is the same ring ${\cal R}$ albeit expressed in a different way.
An explicit order-by-order expression for $e_1'=e_1+\dots$ may be obtained by solving \eqn{eq:GeneratorEquation}.

\bibliographystyle{JHEP}
\bibliography{references}
\end{document}